\documentclass[twocolumn,floatfix, longbibliography, prx,superscriptaddress]{revtex4}

\usepackage{color}
\usepackage{array}
\usepackage{amsmath}
\usepackage{mathtools}
\usepackage{amssymb}
\usepackage{wasysym}
\usepackage{bm}
\usepackage{bbm}
\usepackage{graphicx}
\usepackage{txfonts}
\usepackage{epsfig}
\usepackage{hyperref}
\usepackage{multirow}

\newcommand{\abs}[1]{|#1|}
\newcommand{\bra}[1]{\langle \, #1 \,|}
\newcommand{\ket}[1]{|\, #1 \, \rangle}
\newcommand{\bket}[2]{\langle \, #1 \,|\, #2 \, \rangle}
\newcommand{\boket}[3]{\langle\, #1 \,|\, #2 \,|\, #3 \,\rangle}
\newcommand{\ketbra}[1]{\ket{#1} \bra{#1}}

\newcommand{\be}{\begin{equation}}
\newcommand{\ee}{\end{equation}}
\newcommand{\bc}{\begin{center}}
\newcommand{\ec}{\end{center}}

\newcommand{\non}{\nonumber}

\begin{document}

\title{Hardware-efficient fermionic simulation with a cavity-QED system}
\author{Guanyu Zhu}
\email{gzhu123@umd.edu}  
\affiliation{Joint Quantum Institute, NIST/University of Maryland, College Park, MD 20742 USA}
\author{Yi\u{g}it Suba\c{s}\i}
\affiliation{Theoretical Division, Los Alamos National Laboratory, Los Alamos, NM 87545, USA}
\author{James D.~Whitfield}
\affiliation{Department of Physics and Astronomy, Dartmouth College, Hanover, NH 03755, USA}
\author{Mohammad Hafezi}
\affiliation{Joint Quantum Institute, NIST/University of Maryland, College Park, MD 20742 USA}
\affiliation{Department of Electrical and Computer Engineering and Institute for Research in Electronics and Applied Physics,
University of Maryland, College Park, MD 20742, USA}
\date{\today}
\begin{abstract}

In digital quantum simulation of fermionic models with qubits, non-local maps for encoding are often encountered. Such maps require linear or logarithmic overhead in circuit depth which could render the simulation useless, for a given decoherence time. Here we show how one can use a cavity-QED system to perform digital quantum simulation of fermionic models.  
	In particular, we show that  highly nonlocal Jordan-Wigner  or Bravyi-Kitaev  transformations can be efficiently implemented through a hardware approach. The key idea is using ancilla cavity modes, which are dispersively coupled to a qubit string, to collectively manipulate and measure qubit states.
	 Our scheme reduces the circuit depth in each Trotter step of the Jordan-Wigner encoding by a factor of $N^2$, comparing to the scheme for a device with only local connectivity, where $N$ is the number of orbitals for a generic two-body Hamiltonian.  Additional analysis for the Fermi-Hubbard model on an $N\times N$ square lattice results in a similar reduction. We also discuss a detailed implementation of our scheme with superconducting qubits and cavities.  
\end{abstract}
\maketitle

\section{Introduction}
Quantum computers are widely touted as a new frontier for simulating quantum systems~\cite{Feynman_simulating_1982,lloyd_universal_1996}.  The simulation of quantum chemistry~\cite{Kassal:2011ed, Kelly:2015jca,  Whitfield:2016vi, Bravyi:2017wb, Kandala:2017wj}, strongly correlated fermionic systems~\cite{Wecker:2015kdb, Dallaire2016a, Dallaire2016b, Bauer:2016fc, Kreula2016}, and lattice gauge theories~\cite{Zohar2016, Zohar2017},  are among the crucial applications~\cite{buluta_quantum_2009}. However, apart from ultracold fermionic atoms, all quantum simulation platforms are based on bosonic/spin degree of freedom.  Therefore, one has to encode the fermionic problem into simulation-friendly spin models. 

In the literature, there are a number of methods for doing so and we will focus on the methods that require implementing a non-local map, e.g., Jordan-Wigner (JW) or Bravyi-Kitaev (BK) mappings~\cite{Bravyi:2002cf}.  Our approach relies on the use of a cavity-QED system to achieve the non-local coupling directly.  This is in contrast to other ideas for improving the non-locality of the fermion-spin mapping such as direct simplification of the quantum circuit~\cite{Hastings:2014uj} or using gate teleportation~\cite{Raeisi:2012hc} to lower the cost of the Jordan-Wigner and Bravyi-Kitaev schemes.  Another alternative to the approach taken here is to introduce additional qubits to achieve improved locality of the spin-representations of fermonic operators ~\cite{Verstraete:2005,Ball:2005,Whitfield:2016vi}.     Lastly, we mention a recently introduced technique for quantum simulation using plane waves rather than typical electronic structure basis sets composed of quasi-local Gaussian orbitals~\cite{Babbush17}. The approach taken there has been show to achieve linear circuit depth for a certain class of electronic systems.  We do not pursue subspace encodings and consider arbitary electronic systems with a focus on approaches that directly implement the non-local maps rather than circumventing them.


Here, we present a hardware-efficient scheme to perform digital fermionic simulations on a physical system made of spins. Our approach makes use of cavity-QED physics \cite{birnbaum_photon_2005, Jiang:2008gs, Douglas:2015hda, lezTudela:2015gd}, where one or several ancilla cavity modes are used to encode, simulate the Hamiltonian and measure the desired observables.  The selective non-local coupling of ancillae to a qubit string allows for implementation of  JW and BK mappings in one shot and reduces the simulation time. More specifically, in exponentiating each term of the Hamiltonian, our scheme reduces the circuit depth of both JW and BK to $O(1)$ operations. 
This improvement reduces the simulation time, and therefore, mitigates the decoherence effects. 

We then present an experimental implementation of our scheme in a circuit-QED platform \cite{Blais:2007hh, Schoelkopf:2008vi,  houck2012, Hoffman:2011fz, Nigg:2012jj, Schmidt:2013us,  Barends:2014fu,  Raftery:2014jk, Chiesa:2015vo, HacohenGourgy:2015th, Dalmonte:2015kj, Fitzpatrick:2017cv, Lin:2017uz}, where experimental progress on fermionic and quantum chemistry simulation has been recently achieved \cite{Kelly:2015jca, Kandala:2017wj}. In particular, we use dispersive coupling of microwave cavity photons to superconducting qubits \cite{Nigg:2012jj, Lin:2017uz} to generate non-local string operations non-perturbatively.  This digital approach offers  better scaling in the collective gate time than a previous analog scheme where multi-spin interactions are generated perturbatively \cite{Paik:2016ki}, resulting in an exponential decrease with the number of Pauli operators to be implemented. 
Moreover, experimental advances have been achieved in probing inhomogeneity in resonate frequencies in the context of both superconducting qubit-array and resonator-lattice \cite{Neill:2017wy, Ma:2016hh}, and hence pave the way for the realization of collective many-body gates.
Therefore, our scheme is preferable for implementing large strings, and it also remedies the disadvantage of circuit-QED architecture, i.e.~low connectivity, compared to ion trap architectures \cite{Linke:2017bz}.   

Furthermore, we compare our scheme to conventional local schemes for various fermionic models, such as Fermi-Hubbard model and generic Coulomb Hamiltonian.  In these comparisons, we introduce a parallelization scheme which further improves the simulation. Specifically, by parametrically coupling multiple cavity modes, we further decrease the circuit depth for each Trotter step by an additional factor of $N$.  This results in an overall $O(N^2)$ reduction for Jordan-Wigner and Bravyi-Kitaev transformation in the cases of a Fermi-Hubbard model on an $N$-by-$N$ lattice and a quantum chemistry problem with $N$ orbitals, implemented on a device with local connectivity.

\section{Results}

\subsection{Fermionic encoding with the non-local cavity-QED interaction.}

\subsubsection{Coulomb Hamiltonian and Fermionic encoding}
We consider a generic electronic model with hopping and 2-body Coulomb interaction.   The form of the Hamiltonian is given by
\be\label{CoulombHamiltonian}
H=   \sum_{i,j} \kappa_{ij} (c_i^\dag c_j + \text{H.c.}) +   \sum_{i, j, k, l} V_{ijkl} c_i^\dag c_j^\dag c_k c_l.
\ee
Here, $\kappa_{ij}$ is the hopping matrix and $V_{ijkl}$ represents the interaction matrix.    The indices $i,j,k$ and $l$ can label orbitals either in real-space or the reciprocal-space and can also absorb spin indices.

\begin{figure}
\centering
\includegraphics[width=0.8\columnwidth]{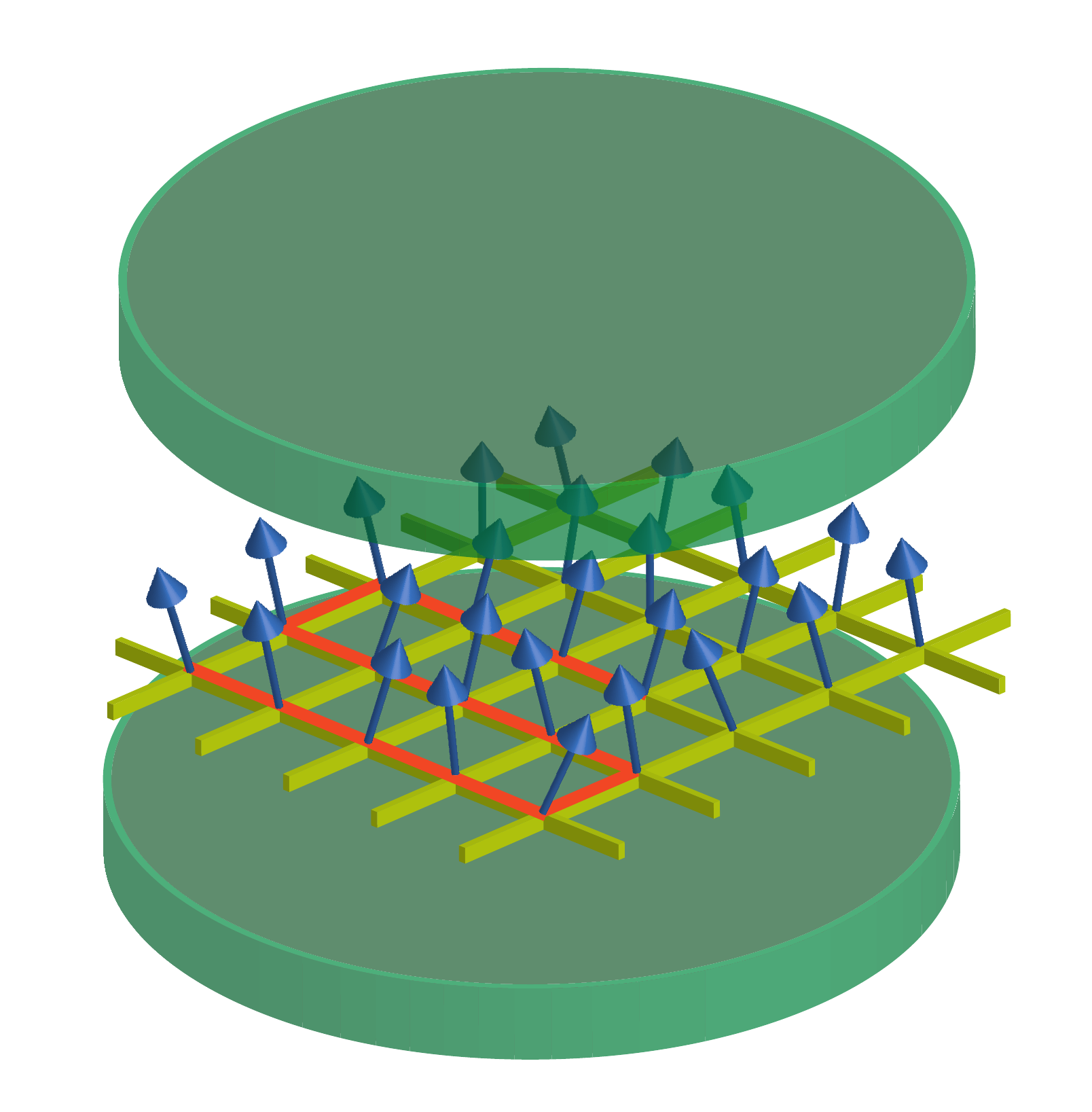}
\caption{Conditional string operation realized in a cavity-QED system. The Jordan-Wigner string (red) in the 2D qubit lattice can be chosen as a snake shape.}
\label{fig:cavity-QED}
\end{figure}

In order to simulate fermions with qubits, the simplest scheme is the Jordan-Wigner transformation:
\be
c_j =\sigma^+_j \prod_{j'<j} \sigma^z_{j'}, \quad c^\dag_j =\sigma^-_j \prod_{j'<j} \sigma^z_{j'},
\ee
The index $j$ can be used to label sites in any dimension.  For example, the string in 2D can be chosen as a `self-avoiding snake' as illustrated by the red string in Fig.~\ref{fig:cavity-QED}. In addition to the JW transformation, the Bravyi-Kitaev transformation~\cite{Bravyi:2002cf} also requires strings of Pauli operators although the form is more complicated (see Appendix VI). The length of Pauli strings are on average logarithmically shorter than JW using the Bravyi-Kitaev transformation.    
In order to implement the time evolution with such string operators, we will consider using the cavity-assisted conditional  string operation in the following sections.  

\subsubsection{Cavity-QED interaction and controlled-string operation}

We consider the quantum non-demolition (QND) interaction \cite{scully1999quantum} of a cavity-QED system in the dispersive regime:
\be\label{QND}
H_\text{QND} = \chi a^\dag a \sum_j \sigma^z_j,
\ee
where $\chi$ is the dispersive interaction strength.

We prepare the cavity photon state in the restricted subspace $n_a = 0,1$.   For circuit-QED implementation, the cavity nonlinearity introduced by the qubits are large enough, such that the cavity itself can be operated as a qubit.  To collectively manipulate a qubit string, we simply apply the dispersive interaction for a period of $\tau$. The time evolution operator is expressed as
\begin{align}
U(\tau)= \bigg[\prod_j (\cos (\chi \tau) - i \sin (\chi \tau) \sigma^z_j)\bigg]^{n_a}.
\end{align}
Here, we used the property that photon and spin operators commute, and the Pauli-matrix property $(\sigma^z_j)^2=1$.  If we choose the operation time to be $\tau = \pi / (2\chi)$, we end up with
\begin{align}\label{controlZ}
U\bigg(\frac{\pi}{2\chi}\bigg) = \mathbbm{1}_q \otimes \ketbra{0}_a + (-i)^N \prod_j \sigma^z_j \otimes \ketbra{1}_a.
\end{align}
 The additional phase factor $(-i)^N$ depends on the length of the string and can be cancelled by applying an additional phase gate on the ancilla cavity, and we call the resulting evolution operator $C_{\overline{Z}}$, i.e., a conditional-$\overline{Z}$ string operator, controlled by the cavity photon state: (1) If $n_a=0$, no operation is performed; (2) If $n_a=1$, a string operator $\overline{Z}=\prod_j \sigma^z_j$ is applied.  Such a cavity-controlled string operation has also been proposed to manipulate and engineer the topological ground state of the toric-code model \cite{Jiang:2008gs, Muller:2011kz, Mezzacapo:2014ww}.

\begin{figure*}
\centering
\includegraphics[width=2\columnwidth]{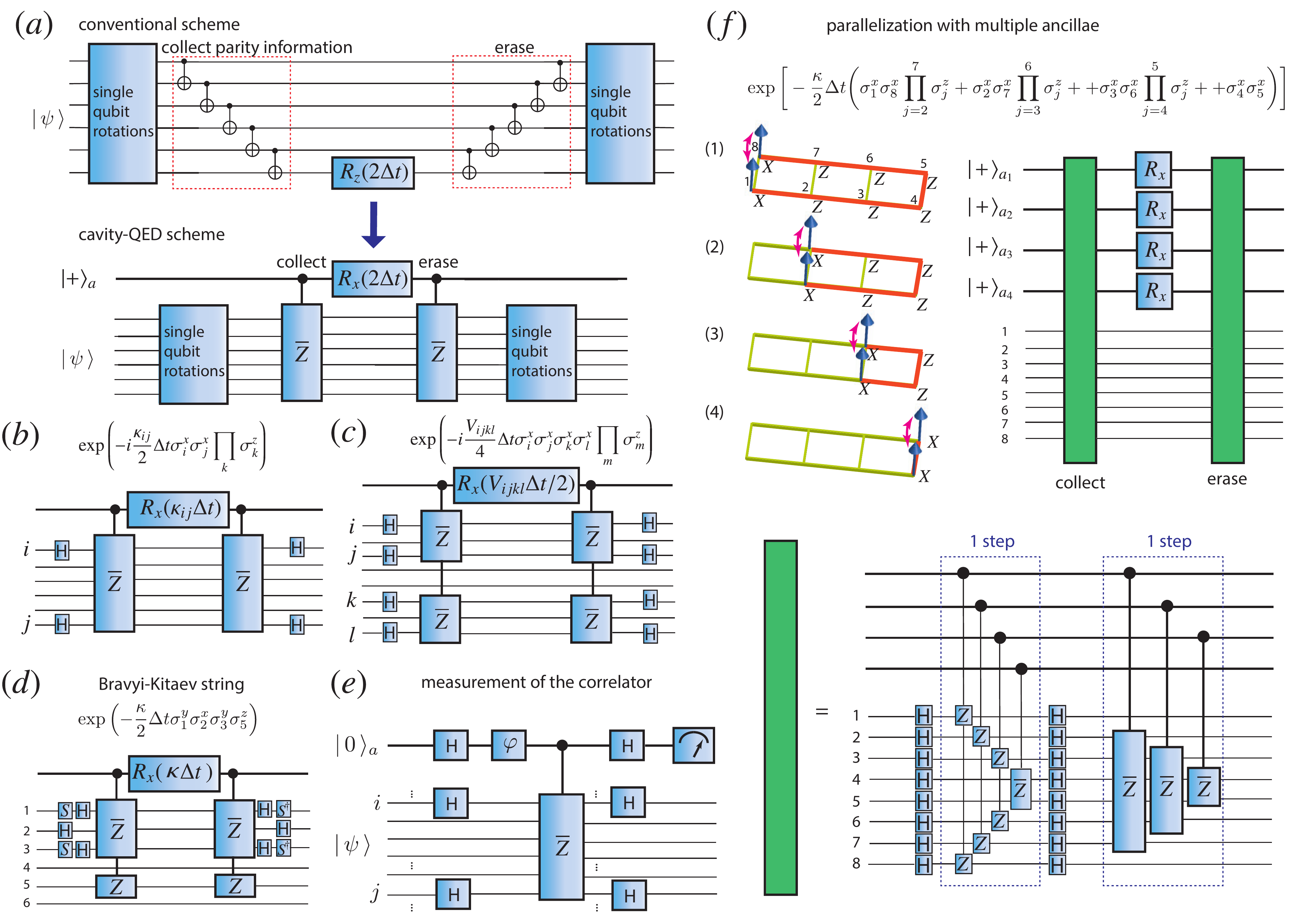}
\caption{(a) Arbitrary string operator exponentiated with conventional approach using a CNOT ladder to collect the parity information.  The whole process can be performed collectively using cavity-QED approach with conditional string operation to realize the exponentiation of the string operator, which reduces the number of gates and the circuit depth by a factor of $1/N$. (b) Exponentiation of a hopping sub-term with the action of pairs of Hadamard gates on sites $i$ and $j$.  (c) Exponentiation of an interaction sub-term with the action of pairs of Hadamards on site $i$, $j$, $k$ and $l$.  (d) Exponentiation of a hopping sub-term in the Bravyi-Kitaev encoding. (e) Measurement of the static correlator $  \langle \psi | \sigma^x_i \sigma^x_j \prod_k \sigma^z_k | \psi \rangle$ with a Hadamard-test circuit.  The expectation value of the correlator can be extracted from the cavity ancilla readout.  (f) Exponentiation of 4 hopping terms in parallel with the coupling to 4 cavity ancillae. In order to switch the ``head" and ``tail" of each string to Pauli-$X$ operator, we split the strings into $\overline{X}$ and $\overline{Z}$ parts.  The $C_{\overline{X}}$ can be implemented with $C_{\overline{Z}}$ sandwiched by parallel Hadamards on the qubits. All the gates in the blue-dashed box are implemented in parallel by multi-mode QND interaction [Eq.~\eqref{multi-QND}].  }
\label{fig:JW-ladder}
\end{figure*}

\subsubsection{Exponentiation of the string operators, time evolution and phase estimation}
In order to perform digital quantum simulation of a Fermionic Hamiltonian $H$, one needs to perform Trotter evolution with small time steps \cite{lloyd_universal_1996}, i.e., $e^{-iH\Delta t}$. After breaking the Hamiltonian down to sub-terms $H=\sum_q h_q$,   
one exponentiates each of these sub-terms as $e^{-ih_q \Delta t}$. The sub-term $h_q$ is composed of a qubit string operator. For example, a hopping term in Eq.~\eqref{CoulombHamiltonian} is represented by qubit operators under JW encoding as $h_{ij}$$=$$\kappa_{ij} (\sigma^+_{i} \sigma^-_{j}+\text{H.c.}) \prod_{k \in \text{string}}\sigma^z_{k}$. This can  be split into two pieces $h_{ij}^{(1)}=\frac{1}{2} \kappa_{ij} \sigma_{i}^x \sigma^x_{j}  \prod_{k \in \text{string}}\sigma^z_{k}$ and $h_{ij}^{(2)}=\frac{1}{2}\kappa_{ij}  \sigma_{i}^y \sigma^y_{j} \prod_{k \in \text{string}}\sigma^z_{k}$, and will be exponentiated separately.   
The conventional approach realizes the exponentiation of these string terms by a CNOT ladder (a sequence of nearest-neighbor CNOTs) illustrated in Fig.~\ref{fig:JW-ladder}(a) (upper panel, see Appendix I for details).  
Here, we present a hardware-efficient quantum circuit which uses the cavity-controlled string operation [Eq.~\eqref{controlZ}] as shown in Fig.~\ref{fig:JW-ladder}(a) (lower panel). The essence is to collect the global parity information into the cavity ancilla with a single $C_{\overline{Z}}$ gate and another $C_{\overline{Z}}$ gate to erase the parity information after the rotation of the ancilla along x-axis by an angle $2\Delta t$.  Note that this circuit reduces the number of gates and circuit depth by a factor of $N$ ($N$ being the length of the string) due to its non-local and highly-parallel feature, and hence greatly reduces the operation time.

To derive the properties of the circuit, we start with the conditional string operation $C_{\overline{Z}}$, and the rotation of the ancilla
\be
R_x(2\Delta t) = \mathbbm{1}_q \otimes e^{-i \Delta t X_a} =\mathbbm{1}_q \otimes [ \cos (\Delta t)  \mathbbm{1}_a  - i \sin (\Delta t) X_a],
\ee
where $X_a$ is the Pauli-X operator of the ancilla photon state.
The three successive gates $C_{\overline{Z}} R_x(2\Delta t)  C_{\overline{Z}} $ can be expressed as
\begin{align}\label{exponentiation}
\nonumber C_{\overline{Z}} R_x(2\Delta t)  C_{\overline{Z}}=  \cos (\Delta t)  \mathbbm{1}_q \otimes  \mathbbm{1}_a - i \sin (\Delta t)\ \overline{Z} \otimes X_a  \\
\quad = (e^{-i \Delta t \ \overline{Z}})^{X_a}=e^{-i \Delta t \overline{Z}} \otimes \ketbra{+}_a + e^{i \Delta t \overline{Z}} \otimes \ketbra{-}_a,
\end{align}
where we have used the property $\overline{Z}^2= \mathbbm{1}_q$.
The final expression represents a conditional evolution with the non-local many-body Hamiltonian $H_\text{string}=\overline{Z}=\prod_{j \in \text{string}} \sigma^z_j$, controlled by the ancilla photon state $\ket{\pm}_a$.

In general, arbitrary many-body interactions along the string can be exponentiated, by choosing the proper single-qubit rotations in the beginning and end of the circuit [see Fig.~\ref{fig:JW-ladder}(a)].   In Fig.~\ref{fig:JW-ladder}(b,c), we show explicitly the circuits to implement the exponentiation of the hopping sub-term $h_{ij}^{(1)}=\frac{1}{2} \kappa_{ij} \sigma_{i}^x \sigma^x_{j}  \prod_{k \in \text{string}}\sigma^z_{k}$ and the interaction sub-term $h_{ijkl}^{(1)}=\frac{1}{4} V_{ijkl} \sigma_{i}^x \sigma^x_{j} \sigma^x_{k} \sigma^x_{l} \prod_{m \in \text{string}}\sigma^z_{m}$ coming from the Coulomb interaction term in Eq.~\eqref{CoulombHamiltonian}, both under JW encoding. Here, we have used Hadamard gates to turn certain $\sigma^z$ operators into $\sigma^x$ with the identity $\text{H}_j \sigma^z_j \text{H}_j =\sigma^x_j $.  On the other hand, a typical term in the Bravyi-Kitaev encoding may involve all types of Pauli operators, e.g., $\sigma^y_1 \sigma^x_2 \sigma^y_3 \sigma^z_5$. This qubit string can be exponentiated with the circuit in Fig.~\ref{fig:JW-ladder}(d), where the combined Hadamards and phase gates ($S$ and $S^\dag$) realized with a single pulse turn the $\sigma^z$ operators into $\sigma^y$.

If one starts the ancilla in the $\ket{+}_a$ ($\ket{-}_a$) state, one only gets forward (backward) evolution after $n$ Trotter steps, $e^{-i n\Delta t H}$ ($e^{i n\Delta t H}$), as suggested by Eq.~\eqref{exponentiation}.
However, if one starts with the ancilla in state $\ket{0}_a = \frac{1}{\sqrt{2}} (\ket{+}_a+\ket{-}_a)$,  one gets a conditional evolution 
$CU$$=$$e^{-iHt} \ketbra{+}_a + e^{iHt} \ketbra{-}_a$, where $t=n\Delta t$. This property can be applied to quantum phase estimation \cite{Kitaev:1995tq, nielsen2000quantum} for extracting energy spectrum and state preparation (see Appendix VIII for details).  
Note, after the state preparation, one can extract fermionic correlation function such as   $C_{ij} = \langle \psi| c^\dag_i c_j |\psi\rangle=\langle \psi|\sigma^+_i \sigma^-_j \prod_k \sigma^z_k |\psi \rangle$ with conditional string operations.  For example, the circuit shown in Fig.~\ref{fig:JW-ladder}(e) implements the xx-part of the correlator, i.e.~$  \langle \psi | \sigma^x_i \sigma^x_j \prod_k \sigma^z_k | \psi \rangle$, where setting $\phi=0$ ($\phi=\pi/2$) in the phase gate gives the real (imaginary) part. The measurement of  dynamical correlator is discussed in Appendix VII.

\subsubsection{Parallelizations with multiple ancillary cavity-modes}\label{sec:parallelization}

Another advantage of the cavity-QED approach is that one can further parallelize the exponentiation of all the mutually commuting sub-terms $h_{ij}$ using multiple cavity ancillae. This can be realized with multiple cavities or different modes in the same cavity as discussed further in the next section. Parallelization is trivial if the string operators to be exponentiated do not overlap with each other.  It is also possible to exponentiate multiple overlapping strings in parallel, namely $\prod_\nu e^{i\kappa \Delta t \overline{\mathcal{S}}_{\nu}}$, where $\nu$ labels different strings.   A concrete example is exponentiating hopping terms between two neighboring rows in parallel which appears in the Hubbard model [illustrated in Fig.~\ref{fig:JW-ladder}(f)]. The detailed derivation can be found in METHODS.

\begin{figure*}
  \includegraphics[width=2\columnwidth]{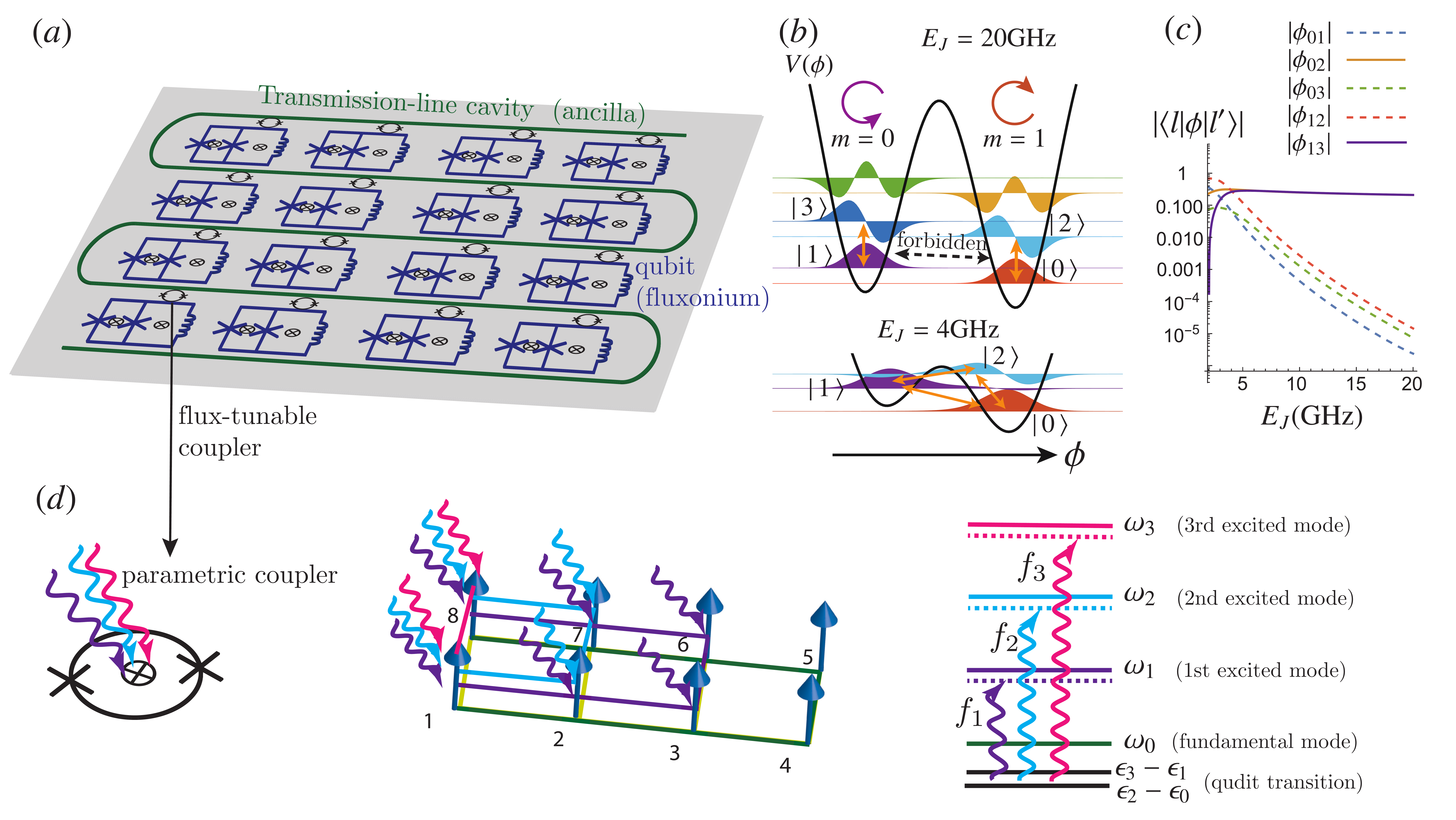}
  \caption{(a) Schematics of a circuit-QED realization: superconducting qubits coupled to a transmission-line cavity with flux-tunable inductive couplers. In particular, we consider using fluxonium circuit as our qubit, and operate it  in the vicinity of half flux quantum into the main loop (the right loop between inductor and junction). (b) The wavefunction is illustrated for $E_C=0.5$ GHz, $E_L=0.75$ GHz, $\Phi_\text{ext} =0.4 \Phi_0$ and tunable $E_J$. For $E_J=20$ GHz (top), the states are trapped deep in the wells  corresponding to persistent-current states flowing in opposite directions  (with winding numbers $m=0$ and $m=1$ respectively). The inter-well transitions are forbidden (dashed arrow), and only intra-well transitions (such as 0-2 and 1-3) are allowed (solid arrows). For $E_J=4$ GHz (bottom), the well is shallow and all transitions are allowed. (c) Magnitudes of phase matrix elements $\abs{\phi_{ll'}}$ as a function of $E_J$ (tunable by external flux through the junction loop on the left). At large $E_J$, $\abs{\phi_{01}}$, $\abs{\phi_{03}}$ and $\abs{\phi_{12}}$ (dashed lines) are exponentially suppressed. The parameters are based on Ref.~\cite{Lin:2017uz}. (d) For further parallelization of multiple terms with overlapping strings, qubits are coupled  to multiple ancillary cavity modes through periodically modulating the couplers with multiple tones.  The qudit transition frequencies $\epsilon_2-\epsilon_0$ and $\epsilon_3-\epsilon_1$  are up-converted close to multiple cavity frequencies $\omega_\nu$ to induce multiple QND interactions \textit{in parallel}.}
\label{fig:circuit_design}
\end{figure*}

\subsection{Implementation with circuit-QED architecture}

In this section, we focus on the experimental implementation of the QND interactions of Eq.~\eqref{QND}.  We also discuss implementation of parallelization with multiple ancilla modes in the same cavity either by higher level contribution or alternatively by periodical modulation of the flux couplers.

\subsubsection{Realization with circuit QED}
We consider a collection of multi-level superconducting qudits inductively coupled to a single or multiple transmission-line cavities or 3D cavities as shown in Fig.~\ref{fig:circuit_design}(a).  The simplest case with one cavity mode can be described by a generalized Tavis-Cummings model \cite{Zhu:2013kf}:
\begin{align}
\nonumber H_\text{cQED}=&H_0 +V, \quad H_0= \omega a^\dag a +  \sum_j\sum_l \epsilon_l \ {_j}\ketbra{l}_j,   \\
V=&  \sum_j \sum_{l,l'} g_{ll'}  {_j}\ket{l}\bra{l'}_j(a+a^\dag).
\end{align}
Here, $a$ is the annihlation operator for the cavity mode with frequency $\omega$, $\ket{l}_j$ represents the $l^\text{th}$ level of the $j^\text{th}$ qudit with corresponding energy $\epsilon_l$, and $g_{ll'}$$=$$g\langle l | \phi | l' \rangle \equiv g\phi_{ll'}$ is  proportional to the inductive coupling strength $g$ and the phase matrix element ($\phi$ being the superconducting phase operator).  The strength $g$ can be made uniform even in the presence of non-uniform mode function with the flux-tunable inductive coupler \cite{Chen:2014cw}, as shown in Fig.~\ref{fig:circuit_design}(a).

In the dispersive regime, namely 
\be\label{dispersive_condition}
\sqrt{N} \abs{g_{ll'}} \ll \abs{\Delta_{ll'}}, \quad  \text{where }\Delta_{ll'}=\epsilon_l- \epsilon_{l'}-\omega,
\ee
 ($N$ represents the total number of coupled qudits and $\Delta_{ll'}$ the detuning), one can adiabatically eliminate the direct inductive coupling $V$ between qudits and the cavity. The effective Hamiltonian after a Schrieffer-Wolff transformation \cite{schrieffer_relation_1966, Zhu:2013kf, Zhu:2013cm} up to second-order is given by
\begin{align}
\nonumber H_\text{eff}=& H_0 + \sum_{j,l} \chi_l \ a^\dag a \ {_j}\ketbra{l}_j + \sum_{j,l} \kappa_l \ {_j}\ketbra{l}_j \\
&+ \sum_{j \neq j'}\sum_{l \neq l'} \mu_{ll'} \ {_j}\ket{l} \bra{l'}_{j'}  +O(g^4).
\end{align}
Apart from $H_0$, the terms appearing in second-order perturbation have three types: (1) The energy shift of level $l$ is given by: $ \displaystyle  \chi_l = \sum_{l' \neq l} \chi_{ll'} =\sum_{l' \neq l}  g^2_{ll'} \left(\frac{1}{\Delta_{ll'}}-\frac{1}{\Delta_{l'l}}\right)$, summed over the contributions $\chi_{ll'}$ from virtual transitions to all other levels $l'$, where the first term is AC Stark and the second term is Bloch-Siegert shift, in the absence of rotating-wave-approximation; (2) the Lamb shift $\displaystyle \kappa_l =  \sum_{l' \neq l} \frac{g^2_{ll'}}{\Delta_{ll'}} $ which only renormalizes the qudit energy level: $\epsilon_l \rightarrow \epsilon_l + \kappa_l $; (3) the flip-flop interactions between any two qudits mediated by virtual photons with strength $\displaystyle  \mu_{ll'} = \sum_{l'' \neq l, l'} \frac{g_{ll''}g_{l''l'}}{2} \bigg(\frac{1}{\Delta_{ll'}}-\frac{1}{\Delta_{l''l}}+\frac{1}{\Delta_{l'l''}}-\frac{1}{\Delta_{l''l'}}\bigg)$, which we need to cancel out to avoid the induced cross-talk errors in our many-body gates.
One can choose specific superconducting circuits, such as fluxonium \cite{manucharyan_fluxonium:_2009, Zhu:2013kf, Zhu:2013fa, Lin:2017uz} focused here (alternatively flux qubit \cite{chiorescu_coherent_2003} or protected 0-$\pi$ qubit \cite{Brooks:2013hi, Dempster:2014tg}). In particular, we consider the situation that phase matrix elements obtain selection-rule property \cite{Zhu:2013fa, Lin:2017uz, Earnest:2017tf} at large ratio of Josephson and charging energy $E_J/E_C$ (e.g.~$E_J$$=$$20$ GHz, with fixed $E_C$$=$$0.5$ GHz from now on): $ \phi_{01}$$=$$\phi_{12}$$=$$\phi_{03}$$=$$0$ as shown in Fig.~\ref{fig:circuit_design}(c).  In the case of fluxonium, this is due to the feature that the ground and excited states are persistent-current states with different winding numbers $m$, which can be seen from their wavefunctions being trapped in different wells of the Josephson potential $-E_J\cos{\phi}$ and have negligible overlap  [Fig.~\ref{fig:circuit_design}(b)]. Therefore, the contribution from $\chi_{01}$ (as well as any other inter-well virtual transition) is nearly zero ($<10^{-5}$ at $E_J$$=$$20$ GHz).  
A QND interaction $H_\text{QND}= \sum_{j} \chi a^\dag a \sigma^z_j$ arises in second-order perturbation with strength $\chi=\sum_{l}(\chi_{0l}-\chi_{1l})/2$, while the nonzero contributions are from intra-well virtual transitions to higher levels, such as $\chi_{02}$ and $\chi_{13}$, which has recently been experimentally observed (see Ref.~\cite{Lin:2017uz}).
On the other hand, the single-excitation flip-flop term ${_j}\ket{0}\bra{1}_{j'}$ disappears ($\mu_{01}$$=$$0$) due to the forbidden inter-well transitions ($g_{01}$$=$$g_{12}$$=$$g_{03}$$=$$0$, etc.), and the lowest-level contribution is from ${_j}\ket{0}\bra{2}_{j'}$.   During the simulation process, we only occupy levels $0$ and $1$ which act as the qubit degree of freedom, therefore the flip-flop process does not play any role and hence will not introduce the unwanted cross-talk error in the many-body $C\overline{Z}$ gate.   When we need to implement single-qubit Hadamard (H) and phase (S) gates to get Pauli-X and Y [Fig.~\ref{fig:JW-ladder}(a)], we can go to the small-$E_J/E_C$ regime (e.g. $E_J$$=$$4$ GHz) by quasi-adiabatically tuning the flux into the junction loop. In this regime, 0-1 transition can be implemented indirectly via a Raman process (0$\rightarrow$2$\rightarrow$1) utilizing the low-lying $\Lambda$-structure \cite{Earnest:2017tf},  as shown in Fig.~\ref{fig:circuit_design}(b, c). A direct transition is also possible since the 0-1 matrix element is sizable and can be accessed by the classical drive.  Alternatively one can stay constantly at an intermediate parameter regime (such as $E_J$$=$$10$ GHz) so that selection rules hold while the suppressed but still non-vanishing 0-1 transition is enabled by enhancing the power of the classical drive.

Note that due to the condition of dispersive regime [Eq.~\eqref{dispersive_condition}], the QND interaction strength $\chi$ has to decrease when the number of coupled qubits $N$ increases due to resonance enhancement.  According to the constraint $g/\Delta \ll 1/\sqrt{N}$ ($\Delta \equiv \text{Min}\abs{\Delta_{ij}}$), one can fix $g$ and increase the detuning magnitude $\abs{\Delta}$ and get the asymptotic scaling $\chi =g\cdot(g/\Delta) \ll g^2/\sqrt{N}$. This scaling is exponentially better than a previous scheme where multi-spin interactions are generated perturbatively \cite{Paik:2016ki} with exponential decreasing interaction strength with the length of the string, i.e., $O(g^N/\abs{\Delta}^{N-1})$.  

For small $N$ [i.e.~$ O(10)$], it is possible to remedy the insignificant decay of maximum interaction strength due to resonance enhancement by varying the parameters (external flux or $E_J$) of individual fluxoniums such that frequency of different qudits ($\epsilon_{l,j}$) are detuned.  The QND interaction strength $\chi$ will not decrease significantly because it contains contributions from multiple levels $\chi_{0l}$ and $\chi_{1l}$. One can then avoid the asymptotic $1/\sqrt{N}$ scaling by modular construction of multiple cavities with $N\sim O(10)$ qubits together connected with quantum teleportation as discussed in Appendix IX. Alternatively, instead of obtaining the QND interaction perturbatively as the above scheme, it is in principle possible to directly engineer the QND (cross-Kerr) interaction such as utilizing nonlinear coupling with Josephson junctions \cite{Nigg:2012jj}. 

Although we focus on fluxonium qubits here, one can generate QND interaction in more general cases for other qubits such as transmons.  In those cases, one can detune the qubit frequency to avoid unwanted flip-flop interactions [for $N \sim O(10)$], or using a balance cavity mode as discussed further in Appendix III.

\subsubsection{Coupling to multiple ancillary modes with parametric coupler}

In order to gain further parallelizability and shorten the time complexity, one can couple the qubits to multiple ancillary cavity modes as mentioned in the previous section, which certainly poses additional experimental challenges. One first needs to selectively address the qubits on different strings with a certain cavity mode which is usually distributed extensively and touches all the qubits. Second, one needs to couple the qubits dispersively to cavity modes with different frequencies.    These two challenges can be solved by one trick, i.e., parametrically modulating the coupling of the qubits to the transmission-line cavity.  One option is to periodically modulate the flux in the inductive coupler shown above in Fig.~\ref{fig:circuit_design}(b) (see e.g.~Refs.~\cite{Roushan:2017kn, Ma:2017ta}) with multiple tones, i.e.~$g_j\big[\Phi(t)\big]$$=$$\sum_{\nu}\tilde{g}_{\nu, j} \cos(f_\nu t)$, where $j$ labels the qubit and $f_{\nu}$ represents the modulating frequencies, with $f_0=0$ (static coupling). The scheme is illustrated in Fig.~\ref{fig:circuit_design}(d). 

The multi-tone modulation technique is mature in microwave-engineering and turns out to be a valuable computational resource. The weight $\tilde{g}_{\nu,j}'$  and driving tones $f_\nu$ are controllable.  We choose $f_\nu$ such that the qubit frequency $\epsilon$ is up-converted to a frequency close to but still off-resonant with the sideband ancillary tones ($f_{\nu}$). In this case, they are dispersively coupled by the QND interaction $H_\text{QND}$$=$$\sum_\nu \sum_j \tilde{\chi}_{\nu,j} a^\dag_{\nu} a_{\nu} \sigma^z_j $ with strength $\tilde{\chi}_{\nu,j} = (\tilde{\chi}_{02}^{\nu,j}-\tilde{\chi}_{13}^{\nu,j})/2$, where $\tilde{\chi}_{ll'}^{\nu,j}= \tilde{g}_{\nu,j}^2/(\epsilon_l -\epsilon_{l'}-\omega_\nu+f_\nu) $. Note that $f_\nu$ can decrease the detuning to make the interaction sizable. We choose $\tilde{g}_{\nu,j}$ such that each qubit is only coupled to the tones of the selected strings, as illustrated in Fig.~\ref{fig:circuit_design}(d) with multiple colors. As we see,  the inductive couplings of qubits 4 and 5 are constant such that the qubits are only dispersively coupled to the fundamental mode $a_0$, while  the couplings of qubits 1 and 8 are modulated by three tones and hence connect the qubits to four cavity modes etc..  
It is clear that the number of cavity modes one can up-convert (or down-convert) to is limited since the up-converted detuning has to be made different to avoid cross-talking between different ancillae modes, but one should be able to couple 10-20 modes. To couple more ancillae, the solution is again teleportation-based modular architecture discussed in Appendix IX.  As we will discuss in the following section, for a Fermi-Hubbard model on a $N \times N$ square lattice in real space, the number of modes one needs to couple to is $N$.   Therefore, for a 100-qubit system which can be realized in the near future for a short-circuit algorithm still requiring no quantum error correction, it is possible to realize our parallelization scheme.

\begin{figure}
\centering
\includegraphics[width=1\columnwidth]{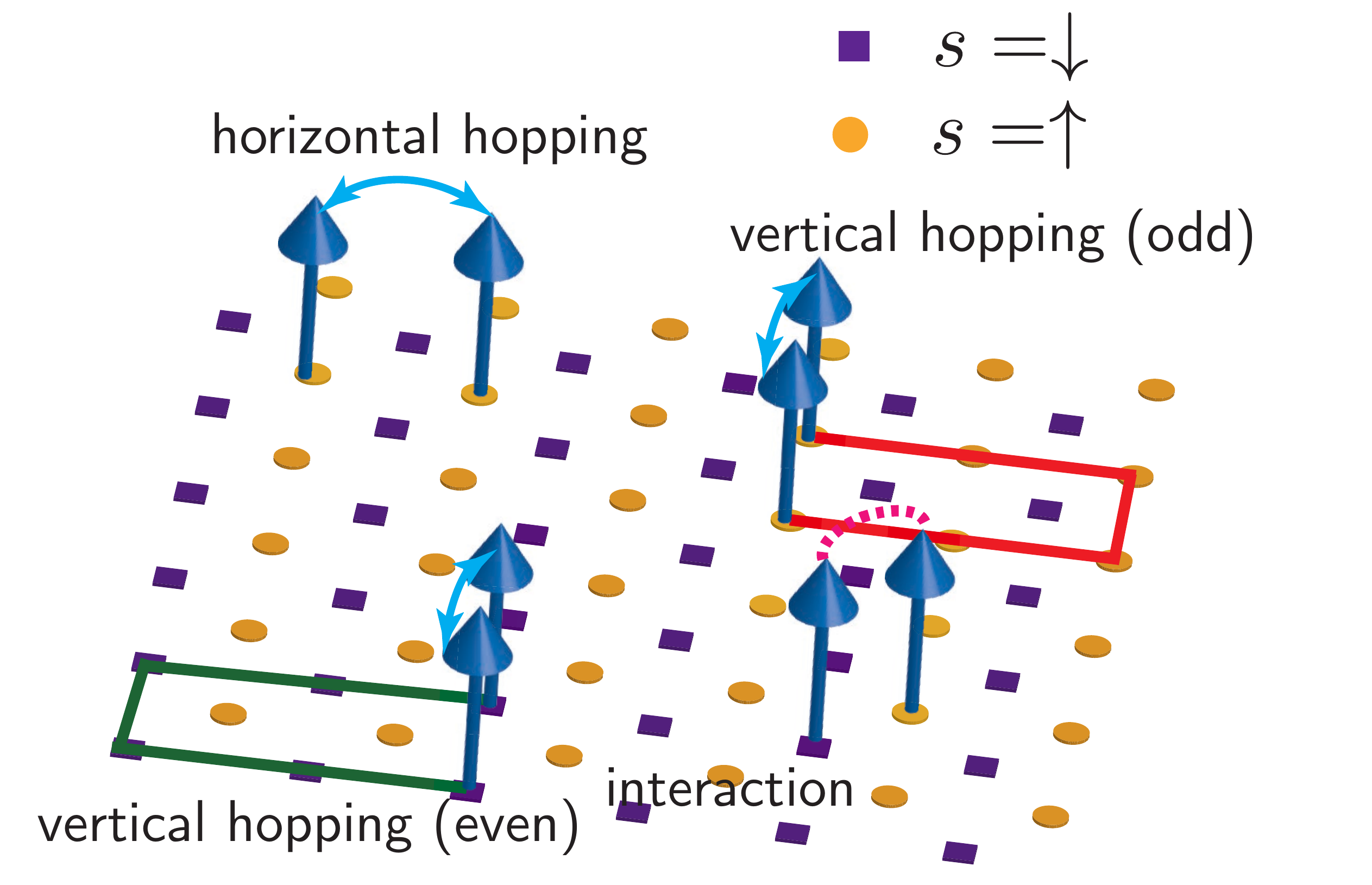}
\caption{Types of terms and Jordan-Wigner strings in a 2D spinful Fermi-Hubbard model on an $N \times N$ lattice.  One can consider it as a checkerboard lattice with two sub-lattices (purple and yellow) representing two spin species ($\downarrow$ and $\uparrow$) respectively. The `even' and `odd' vertical hoppings differs by the location of the strings, which are on the left and right sides respectively.}
\label{fig:jordan-wigner}
\end{figure}

\subsection{Time complexity}\label{sec:complexity}

In the previous sections, we focused on how to exponentiate a single term $h_p$ in the system Hamiltonian $H$$=$$\sum_p h_p$.   In the following, we compare the time complexity (circuit depth) of our cavity-QED approach with the conventional approach of a single Trotter step $e^{-iH\Delta t}$.  

\subsubsection{Fermi-Hubbard model}\label{sec:Hubbard}
As the first example, we consider the  spinful 2D Fermi-Hubbard model in real-space and on an $N \times N$ square lattice.  We use qubits on two sub-lattices to encode fermions with different spin $s = \downarrow$ (purple) or $s=\uparrow$ (yellow) as shown in Fig.~\ref{fig:jordan-wigner}.   The spinful Fermi-Hubbard model is a restricted form of Eq.~\eqref{CoulombHamiltonian} given by
\be\label{SpinfulFermiHubbard}
H_\text{Hubbard}= - \kappa \sum_{\langle i,j \rangle, s} (c_{i, s}^\dag c_{j, s} + \text{H.c.}) +  U \sum_j  n_{j,\uparrow} n_{j, \downarrow},
\ee
where $j\rightarrow(n_x, n_y)$ is a two-component label for the 2D sub-lattice. The first and second terms represent hoppings and on-site Hubbard interaction respectively. 
 The types of terms and their corresponding time complexity is listed below (for more details see Appendix V). 

(1 and 2) \textit{On-site Hubbard interaction} and \textit{Horizontal hopping}:  
translates to ZZ interaction and 2-local flip-flop interaction without string in the qubit representation, both of which have $O(1)$ circuit-depth.   
 (3) \textit{Vertical hopping (even and odd)}:  typically contains a ``snake-shape" JW string (Fig.~\ref{fig:jordan-wigner}) and hence dominates the time complexity.

With one transmission-line cavity coupled to each pair of rows, one can parallelize the vertical hopping terms (see Appendix V for details).
For the vertical hopping between the same pair of rows, one can exponentiate these terms \textit{in series}, resulting in the Trotter step circuit depth (time complexity) $O(N)$. With the multi-mode scheme shown in Fig.~\ref{fig:JW-ladder}(f) and Fig.~\ref{fig:circuit_design}(d), one can exponentiate these terms and reduce the depth to $O(1)$.  In contrast, the conventional approach needs $O(N^2)$ due to the linear overhead of implementing the CNOT ladder in Table~\ref{table:molecule_table}.

\begin{table*}
\begin{tabular}{lllllll}
\hline  
Molecule 
 & \begin{tabular}{l}$\text{BeH}_2$\\ (6 qubits)\end{tabular}\phantom{spc} 
 & \begin{tabular}{l}$\text{BeH}_2$\\ (14 qubits)\end{tabular} \phantom{spc} 
 & \begin{tabular}{l}$\text{H}_2\text{O}$\\ (14 qubits) \end{tabular}\phantom{spc} 
 & \begin{tabular}{l} HCl\\ (20 qubits) \end{tabular}\phantom{spc}
 &  \begin{tabular}{l}LiH\\ (12 qubits)\end{tabular}\phantom{spc}
 & \begin{tabular}{l}$\text{NH}_3$\\ (16 qubits)\end{tabular} \phantom{spc} \\
\hline                                                                                                                                                      
Hamiltonian Pauli terms   &                     164                       & 1150                       & 1858                             & 4427            & 631          & 4973 \\ 
Number commuting groups &     8                         & 43                         & 70                               & 162             & 18            & 178 \\ 
Terms per group &   20.5                      & 26.7                       & 26.5                             & 27.3            & 35.1         & 27.9  \\
Hamiltonian op. weight  &     3.5                       & 6.2                        & 6.2                              & 7.7             & 5.1          & 6.7  \\ 
Average qubit participation  & 12.1                      & 11.8                       & 11.7                             & 10.6          & 15.1             & 11.8      \\            
\hline  
\hline        
\end{tabular}
\caption{\textbf{Summary of various properties of six different molecules (operator information based on Ref.~\cite{Kandala:2017wj, Bravyi:2017wb}).} The first row lists the number of Pauli terms in the Hamiltonian which can be grouped into sets of mutually commuting groups.  The minimum number of such groups and the average number of terms per group appear in rows two and three, which dictate the minimum Trotter-step circuit depth and number of cavity ancillary modes needed for parallelization.  Row four contains the average number Pauli operators in each term which determines the cavity load, i.e.,~ the number of qubits interacting with a single cavity mode simultaneously.  Finally the last row lists the average number of terms within each mutually commuting group that each qubit participates in, which determines the qubit load, i.e.~the number of cavity modes interacting with each qubit simultaneously. }
\label{table:molecule_table}
\end{table*}

\begin{table*}
\begin{tabular}{llllll}
 \hline
 \multicolumn{2}{c}{} & \multicolumn{2}{c}{Conventional local approach \cite{Wecker:2014bm}}\phantom{space}  & \multicolumn{2}{c}{Proposed cavity-QED approach } \\
 \cline{3-6}
  \multicolumn{2}{c}{} & Jordan-Wigner\phantom{spc} & Bravyi-Kitaev & Jordan-Wigner\phantom{spc} & Bravyi-Kitaev \\
  \hline
  \multicolumn{2}{c}{Interaction type} &   \multicolumn{2}{c}{ $g'^2\eta^{-1} \sum_{\langle i,j\rangle} \sigma^z_i \sigma^z_j$ \! \! \! \! \! \! \! \!  \! \! \! \! \! \! \! \! \! \! \! \! \! \! \! \! \! \!}    & \multicolumn{2}{c}{$\chi \sum_j a^\dag a \sigma^z_j $}  \\
\multicolumn{2}{c}{Gate time (ns)}   &\multicolumn{2}{c}{ 40 \! \! \! \! \! \! \! \!  \! \! \! \! \! \! \! \! \! \! \! \! \! \! \! \! \! \! } &   40$\sqrt{N}$  &  $40\sqrt{\log N}$ \\
\multicolumn{2}{c}{Circuit depth to exponentiate a single term}  & $O(N)$    &$O(\log N)$&   $O(1)$ & $O(1)$  \\
\multicolumn{2}{c}{Pulse fidelity of gate control}  & $O(F^N)$    &$O(F^{\log N})$&   $O(F')$ & $O(F')$  \\
\hline \\[2ex]
 \hline

\multicolumn{6}{c}{I. 2D Fermi Hubbard  model in real space ($N \times N$ square lattice)} \\
 \hline
 Trotter step & circuit depth / time complexity  &\multirow{2}{*}{$O(N^2)$}  & \multirow{2}{*}{$ O(N \log N) $} &  $O(N)$, series & $O(N)$, series\\
\cline{5-6}
&    & &  &  $O(1)$, parallel & $ O(1)$, parallel\\\hline\\[2ex]
 \hline
\multicolumn{6}{c}{II. Generic Coulomb Hamiltonian ($N$ orbitals)} \\
\hline 
Trotter step & circuit depth / time complexity  &\multirow{2}{*}{$O(N^5)$}   & \multirow{2}{*}{$O(N^4 \log N)$} &  $O(N^4)$, series  & $ O(N^4)$, series \\
&  &  &  &  $O(N^3)$, parallel  & $ O(N^3)$, parallel \\
\hline
\hline
\end{tabular}
\caption{\textbf{Comparison of the conventional (local) and cavity-QED approaches} with Jordan-Wigner and Bravyi-Kitaev encodings. The interaction strength and gate time are listed. Gate times and interaction strengths are approximate, and are based on Ref.~\cite{Nigg:2012jj} and \cite{Barends:2014fu}.  For the pulse fidelity of gate control, we assume a single pulse has a fidelity $F$ for the qubit control and $F'$ for the cavity control.  Note that the scaling for Bravyi-Kitaev encoding  listed in this table assumes a non-local cavity ancilla which can selective address an arbitrary cluster of connected or disconnected qubits, and in the BK case the number of qubits in the cluter is $O(\log N)$. This is different from the case of a device with only local connectivity where the scaling is essentially the same as the Jordan-Wigner encoding.}
\label{table:comparison_table}
\end{table*}

\subsubsection{The generic Coulomb Hamiltonian}

For the generic Coulomb Hamiltonian described in Eq.~\eqref{CoulombHamiltonian}, which is the relevant model for quantum chemistry or strongly-correlated electronic materials simulated in reciprocal space, the indices $i,j,k$ and $l$ are typically not neighbors.   The type of terms that dominate the computational resource is the 4-local interaction term $V_{ijkl} c_i^\dag c_j^\dag c_k c_l$, which requires a sequence of $O(N^4)$ unitary transformations for a system with $N$ orbitals ($i,j,k,l=1,2,\cdots N$) in a single Trotter evolution step due to all possible choices of the four fermion indices.    Taking into account the JW string, which has length of $O(N)$, the Trotter step circuit depth of the conventional approach becomes $O(N^5)$ \cite{Wecker:2014bm}.

For our cavity-QED approach, we list the circuit depth for the two approaches. (1) \textit{Series}: $O(N^4)$, due to the reduction of the linear overhead of the Jordan-Wigner string.  (2) \textit{Parallel}:  $O(N^3)$, assuming $N$ ancilla cavity modes. The remaining $O(N^3)$ terms cannot be exponentiated in parallel because they do not commute with each other (e.g.~when the first index $i$ coincide, but the remaining 3 indices $j,k,$ and $l$ are all different).   However, note that for an actual quantum chemistry Hamiltonian, although the total number of terms scales as $O(N^4)$, a large number of integrals vanish between distant orbitals or due to symmetry. The number of non-commuting terms also scales as $O(N^3)$ though similarly sparse. This can be seen from the example molecules discussed in Table \ref{table:molecule_table} (operator information collected from Ref.~\cite{Bravyi:2017wb, Kandala:2017wj}), which has typically only $O(N)$ to less than $O(N^2)$ non-commuting terms (equivalent to the minimum number of commuting groups listed in the table). Therefore, there is a huge potential for parallelization in practice.

\subsubsection{Summary of the comparison between cavity-QED and conventional approaches}
Here, we summarize and compare the various properties of the cavity-QED scheme versus the conventional scheme, as shown in Table \ref{table:comparison_table}.

In order to compare both schemes, we first compare their gate time. With the state-of-the-art technology, the second-order QND interaction strength between qubits and cavity with the form   $\chi \sum_j a^\dag a \sigma^z_j$, can typically reach about 50 - 100 MHz \cite{Nigg:2012jj}, corresponding to gate time of 20 - 40 ns.  On the other hand, the conventional approach needs nearest-neighbor CNOT gates between qubits, coming from the second-order ZZ interaction,  $\frac{4 g'^2}{\eta}\sum_{i,j} \sigma^z_i \sigma^z_j$,  (e.g.~due to the third-level contribution in the context of transmon qubits \cite{dicarlo_demonstration_2009}, where $\eta$ is the nonlinearity of the transmon). The typical strength of the ZZ interaction is around 50 MHz \cite{Barends:2014fu}, corresponding to a gate time of 40 ns. Since both types of interactions are of perturbative nature (up to second order), the gate time in both cases are of the same order of magnitude. The relevant parameters are summarized in Table \ref{table:comparison_table}.   We also include the asymptotic prefactor $\sqrt{N}$  (reduces to $\sqrt{\log N}$ with the Bravyi-Kitaev encoding) of the cavity-QED gate time due to the dispersive regime condition [Eq.~\eqref{dispersive_condition}], which can be remedied by the modular architecture connecting multiple cavities (Appendix IX).    The average number of strings (cavity ancilla modes) a single qubit touches simultaneously is of $O(10)$, so one does not need to worry about cross-talk between the ancillae due to frequency crowding in these cases either.

We emphasize that having a scheme with a shorter operation time in each Trotter step enables more evolution steps within the coherence time of the system, and hence increases the precision of the algorithms, such as phase estimation.   Besides the cavity-QED scheme presented in this paper, there are some other schemes which can reduce the overhead due to the non-local string operator, such as Ref.~\cite{Raeisi:2012hc} and \cite{Hastings:2014uj}. We compare our scheme with theirs in Appendix X.  
  
 Another significant advantage of our scheme over the conventional scheme is the gate fidelity, in particular, the fidelity due to the control pulses.    In the conventional scheme, in order to implement $N$ CNOTs in the CNOT ladder, one has to send $N$ control pulses. Assuming the fidelity is $F$ for each pulse, the overall fidelity due to imperfect pulse becomes $F^N$ as shown in Table \ref{table:comparison_table}.   On the other hand, in the case of our many-body gate, one can actually just use a single control pulse with error $F'$ to detune the cavity frequency.   In this case, the overall fidelity due to imperfect pulse is just $F'$, which does not have an exponential decay.  Therefore, our collective many-body gate has a significant advantage in terms of quantum control and pulse fidelity.

\begin{figure*}
\includegraphics[width=2\columnwidth]{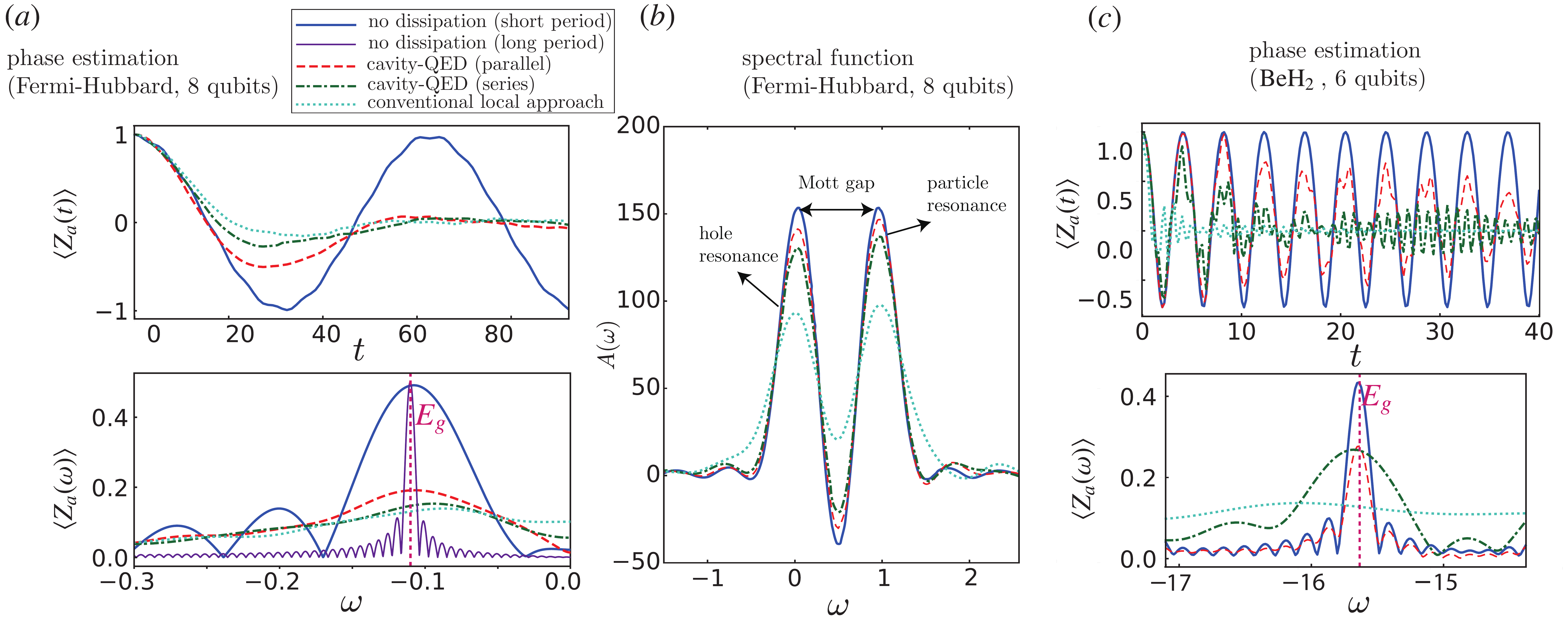}
\caption{Numerical simulation of the measurement protocols for different approaches taking into account dissipation effects (summed over 50 quantum trajectories in each curve), with the following jump operators for qubits and cavity and corresponding decay rate (from Ref.~\cite{Schmidt:2013us}):  $\sigma_j^-$ (10 kHz) , $\sigma_j^+$ (0.05 kHz), $\sigma_j^z$ (50 kHz), $a$ (5 kHz) and $a^\dag$ ( $\sim$ 0 kHz) .  (a) Phase estimation of the 2D Fermi-Hubbard model on a $2 \times 2$ lattice (simulated by 8 qubits), with the parameter: $\kappa=0.1, U=1,$ and 4 electrons in total (half-filling).  The upper panel shows the time-domain signal of the ancilla expectation value, while the lower panel is the Fourier transform of the upper panel in order to extract the ground-state energy. The actual ground-state energy $E_g$ of this model is shown by vertical dashed lines.  Note that all the curves in the lower panel correspond to Fourier transform of the signal in the period $0 \le t \le 100 $,  while the purple curve corresponds to the ideal case with no dissipation and being transformed over a much longer period $0 \le t \le 1000 $ such that the resolution is improved by about 10 times. (b) The spectral function (extracted from the dynamical correlation function) of the Fermi-Hubbard model. The separation between the hole and particle resonance peaks signals the Mott gap.  (c) Phase estimation of the Be$\text{H}_2$ molecule (simulated by 6 qubits).  Due to the signal decay of  the cavity-QED (series) and local approach, we only perform Fourier transform in the period $0 \le t \le  10$. In the numerical simulation, we first subtract all the diagonal terms in the Hamiltonian and then shift it back to recover the eigenenergy, mimicking the actual experimental process in Ref.~\cite{Barends:2014fu}. One can see all but the conventional local approach can locate the ground-state energy $E_g$ (dashed line), while the cavity-QED (parallel) approach has almost a resolution as good as the ideal case with no dissipation, despite the shrink of the peak.}
\label{fig:numerics}
\end{figure*}

\subsection{Numerical simulation in the presence of decoherence}
In this section, we numerically simulate and compare different approaches with two simple but representative experiments: (1) A 2D spinful Fermi-Hubbard model on a $2\times 2$ lattice (simulated by 8 qubits). (2) A quantum chemistry problem, i.e., the outer shell electrons of a Be$\text{H}_2$ molecule (simulated by 6 qubits), which has been simulated with superconducting qubits in a recent experiment \cite{Kandala:2017wj}.

The simulation takes into account decoherence of qubits and cavity, represented by the jump operators
$l_j = \tilde{l}_j\sqrt{\Gamma_j} $, where $\Gamma_j$ is the corresponding decay rate and $\tilde{l}_j$ the normalized operator. 
The  types of jump operators of our numerical simulation is listed in the caption of Fig.~\ref{fig:numerics}, along with the realistic estimation of experimental parameters chosen according to Ref.~\cite{Schmidt:2013us}. 

In particular, we simulate the Kitaev phase estimation protocol (see Appendix VIII) for both systems and for the Fermi-Hubbard model also the measurement of spectral function $A(\omega)$$=$$ -2\text{Im}[G(\omega)]$, where $G(\omega)$ is extracted from the Fourier transform of the dynamical correlators including $\boket{\psi}{c_i(t) c^\dag_j(0)}{\psi}$  (see Appendix VII).  Since both measurement protocols involve time evolution $U(t)$, the dissipation of the system will affect the measurement result, as shown in Fig.~\ref{fig:numerics}. We compare four different situations: the ideal situation without dissipation, the conventional approach, and the cavity-QED approach \textit{in series} and \textit{in parallel} respectively.  Since each approach needs different operation time per Trotter step, the effects of dissipation are different. 

For the Fermi-Hubbard model, we use JW encoding in all cases and three transmission line cavities are needed to couple each pair of rows (four rows in total) \textit{in parallel}. For the Be$\text{H}_2$ molecule, we use the modified BK encoding discussed in Ref.~\cite{Kandala:2017wj}. With this encoding, there are a total of 164 terms, which can be divided into 8 groups, where all the terms in the same group commute with each other, as shown in Table~\ref{table:molecule_table}.  In this case, one can reduce the circuit depth to 8 by exponentiating all the terms in the same group \textit{in parallel} with multiple ancilla modes in the same central cavity. This would require about 20 tones in the flux modulation using the trick in Fig.~\ref{fig:circuit_design}(d).  On the other hand,  the \textit{series} cavity-QED approach will exponentiate all the terms sequentially with a single cavity ancilla.

Regarding to the phase estimation protocol in Fig.~\ref{fig:numerics}(a) and (c), the cavity ancilla expectation $\langle Z_a (t) \rangle$ (Pauli-Z) oscillates in time in the ideal case, i.e. $\langle Z_a (t) \rangle = \cos(E_g t)$, where $E_g$ is the ground-state energy of the prepared eigenstate. Nevertheless, in the presence of decoherence, the signal decays significantly in time, while the peaks in frequency-space signal $\langle Z_a (\omega) \rangle$ also shrinks due to dissipation. For the Fermi-Hubbard model in (a), we prepare the ground state in the beginning, and one can see that $E_g$ (shown by the  dashed line) can be clearly resolved in the biggest peak in $\langle Z_a (\omega) \rangle$ in the blue and purple curves (ideal dissipationless case).  The purple curve has a Fourier transform over the period $0 \le t \le 1000 $, namely 10 times long as the others, and hence has much better resolution. With dissipation, the signal dies out in a short time.  While this peak still has the correct position for the cavity-QED \textit{parallel} approach (red dashed), it shifts slightly for the \textit{series} approach (green dashed) and becomes obscured in the conventional approach local (light blue dashed).   For the phase estimation in Be$\text{H}_2$ molecule in (c), we see that the parallel cavity-QED approach (red dashed) approximates the dissipationless signal (blue) with almost the same resolution of the ground-state energy while the height of the peak is reduced. The \textit{series} cavity-QED approach (green dashed) has significant broadening in the resolution, while the conventional local approach has all the peaks being smeared out and is hence hard to tell the actual energy.

For the spectral function measurement in panel (b) for Fermi-Hubbard model, we prepare the initial state as the ground state.  The two biggest peaks correspond to the hole (left) and particle (right) resonance respectively, and the distance is approximately $U$, namely the Mott gap.   We can see that the dissipation effect leads to the shrinking and asymmetry of the two peaks. The shrinking is proportional to the operation time of different approaches.  The asymmetry is due to the fact that the qubit has much larger loss rate than absorption rate as listed in the figure caption.   Due to our encoding of 0 (1) electron as spin up (down) of the qubit,  the qubit loss induces loss of holes but not particles.   Therefore, the hole peak (left) shrinks more than the particle peak.   In practice, one could choose two different ways of encoding and average the signal to get rid of this asymmetry.

\section{Conclusion and discussion}
In this article, we have shown that, in the context of cavity/circuit-QED architecture, the use of the common cavity modes greatly simplifies the non-local string-like encoding needed for fermionic simulation, such as Jordan-Wigner and Bravyi-Kitaev transforms. In particular, we are able to get rid of a polynomial overhead, i.e., $N^2$  of the Trotter-step circuit depth  in the conventional local approach, which reduces the time complexity of the simulation for a given precision and in turn reduces the decoherence effects.   The non-local quantum control and parallelization of multiple ancilla-controlled processes developed in this  paper may have profound applications in many others areas, such as quantum information processing, lattice gauge theory simulation and measurement of entanglement spectrum in quantum many-body systems \cite{Pichler:2016ec}.

\section{Methods}

\subsection{Derivation of parallelizations with multiple ancillae}

Here, we show the detailed derivation of multi-ancilae parallelization mentioned above. We use conditional string-$\overline{Z}$ operations with multiple cavity ancilla modes, namely
\begin{align}\label{multi-control-Z}
\non C_{\overline{Z}_{\nu}} &= \mathbbm{1}_q \otimes \mathbbm{1}_{a_1} \otimes\mathbbm{1}_{a_2}\otimes\cdots \otimes\ketbra{0}_{a_\nu} \otimes \mathbbm{1}_{a_{\nu+1}} \otimes\cdots \\
 & + \prod_j \sigma^z_{j\in \text{string}(\nu)} \otimes \mathbbm{1}_{a_1} \otimes\mathbbm{1}_{a_2}\otimes\cdots \otimes \ketbra{1}_{a_\nu}\otimes \mathbbm{1}_{a_{\nu+1}} \otimes\cdots,
\end{align}
where each ancilla mode $a_\nu$ is dedicated to a particular string~$\nu$.
This collective gate can be realized by dispersively coupling qubits simultaneously to multiple modes resulting in the QND interaction 
\be\label{multi-QND}
H_\text{QND}'=\sum_\nu \sum_j \tilde{\chi}_{\nu,j} a^\dag_{\nu} a_{\nu} \sigma^z_j.
\ee
As explained below, by proper conditional rotations, we can achieve a generic conditional string-$\overline{\mathcal{S}}$ in different Pauli-bases, i.e. $C_{\overline{\mathcal{S}}_{\nu}}$, where the $\overline{Z}_\nu$ string in Eq.~\eqref{multi-control-Z} is replaced by $\overline{\mathcal{S}}_{\nu}$. 
We consider the case where all the strings commute with each other, i.e.~$[\overline{\mathcal{S}}_\nu, \overline{\mathcal{S}}_{\nu'}]=0$. Thus the conditional-string also commutes, i.e.~$[C_{\overline{\mathcal{S}}_{\nu}}, C_{\overline{\mathcal{S}}_{\nu'}}]=0$. Therefore, following the derivation in Eq.~\eqref{exponentiation},  we can reach the identity
\begin{align}
\non & \prod_{\nu}C_{\overline{\mathcal{S}}_{\nu}} \prod_{\nu'} R^{\nu'}_x(2 \kappa\Delta t) \prod_{\nu''} C_{\overline{\mathcal{S}}_{\nu''}}= \prod_{\nu}C_{\overline{\mathcal{S}}_{\nu}}  R^{\nu}_x(2 \kappa\Delta t) C_{\overline{\mathcal{S}}_{\nu}} \\
=& \prod_{\nu} (e^{i \kappa\Delta t \ \overline{\mathcal{S}}_{\nu}})^{X_{a,\nu}},
\end{align}
where  $R^\nu_x$ and $X_{a,\nu}$ is the x-axis rotation and Pauli-X operator of the ancilla mode $\nu$. If all the ancillae are initiated at $\ket{+}_\nu$, the exponentiation of multiple strings is achieved in parallel, i.e.~$\prod_\nu e^{i \kappa\Delta t \ \overline{\mathcal{S}}_{\nu}}$. 

Now we consider how to convert the conditional-$\overline{Z}$ into conditional-$\overline{\mathcal{S}}$. We illustrate the idea with example shown in Fig.~\ref{fig:JW-ladder}(f)]. This involves turning the head and tail of each string into Pauli-X operators.   To achieve this, we split the $C_{\overline{\mathcal{S}}_\nu}$ operator into two parts applied sequentially (order is arbitrary): the main $C_{\overline{Z}^1_\nu}$ string and the $C_{\overline{X}^2_\nu}$ part in the ends as shown in the green box in Fig.~\ref{fig:JW-ladder}(f). To achieve $C_{\overline{X}^2_\nu}$, we just need to sandwich the $C_{\overline{Z}^2_\nu}$ operators with Hadamards $\text{H}_j$ performed on the qubits in parallel.  The application of all the $C_{\overline{Z}_\nu}$ gates are performed in parallel with multi-mode QND interaction $H'_\text{QND}$ [Eq.~\eqref{multi-QND}].   Therefore, the overall circuit depth of parallelizing $N$ such hopping terms is of  $O(1)$.  The generalization to arbitrary type is shown in Appendix II.

\section{Data Availability}
The data sets generated during and analysed during the current study are available from the corresponding author on reasonable request.

\section{Acknowledgement}
We thank Vladimir Manucharyan for the suggestions of the scheme using fluxonium qubits and providing experimental details and parameters. We thank Ignacio Cirac for pointing out the scaling of dispersive interaction.  We also thank Peter Zoller, Jens Koch and Eran Ginossar for helpful discussions.  GZ and MH were supported by ARO-MURI, NSF-PFC at the JQI, YIP-ONR, and the Sloan Foundation. The work by GZ was performed in part at Aspen Center for Physics, which is supported by National Science Foundation grant PHY-1607611. JDW acknowledges startup funds from Dartmouth College.  This work was performed under the auspices of the U.S. DOE contract No. DE-AC52- 06NA25396 through the LDRD program at LANL.

\section{COMPETING INTERESTS}
The authors declare no competing interests.

\section{Author Contributions}
All authors researched, collated, and wrote this paper.

\begin{appendix}

\section{Derivation and circuit transformation between the CNOT ladder and  the cavity-QED circuit.}\label{sec:circuit_derivation}
Here we derive the analytic expression of the Jordan-Wigner ladder and its variant used in the conventional approach and show how it can be transformed to the cavity-QED circuit we use.

\begin{figure}
  \includegraphics[width=0.7\columnwidth]{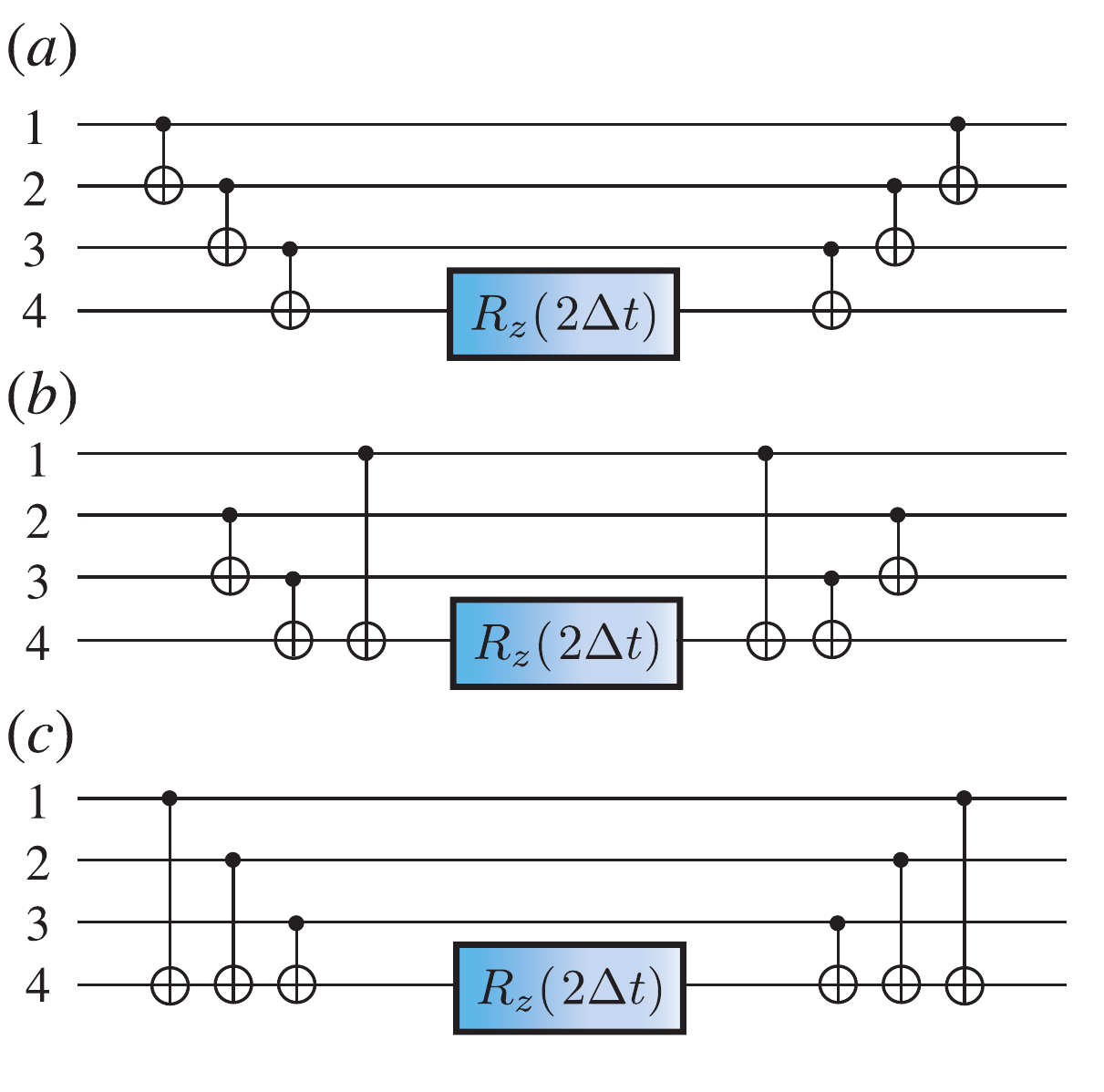}
  \caption{CNOT ladder and its variants (4-qubit version as an illustration).}
\label{fig:circuit-transform}
\end{figure}

We first consider a 4-qubit version of Jordan-Wigner ladder as shown in Fig.~\ref{fig:circuit-transform}(a), the derivation of which can be easily generalized to the $n$-qubit case.   The central derivation relies on the following identity of CNOT gate, i.e.
\be\label{CNOT}
\text{CNOT} (\mathbbm{1} \otimes \sigma^z)  \text{CNOT} = \sigma^z \otimes \sigma^z.
\ee
This identity is essentially a two-qubit basis transformation and tells us that, when acted by the CNOTs on the two sides, the $\sigma^z$ operator of the target qubit grows to a 2-bit $\sigma^z$ string involving the control qubit as well. The quantum circuit for exponentiating a z-string can be described by the following unitary operator:
\be
  U= C_{12} C_{23} C_{34}(\mathbbm{1}\otimes \mathbbm{1} \otimes \mathbbm{1} \otimes e^{i\Delta t \sigma^z_4})C_{34}C_{23}C_{12},    
\ee
where we have abbreviated the CNOT between qubit $i$ and $j$ as $C_{ij}$.  The unitary can be simplified by repetitively using Eq.~\eqref{CNOT} and the identity $\text{CNOT}^2=\mathbbm{1}\otimes \mathbbm{1}$ as follows:
\begin{align}
\nonumber U=& C_{12} C_{23} C_{34}[\mathbbm{1}\otimes \mathbbm{1} \otimes \mathbbm{1} \otimes (\cos (\Delta t) \mathbbm{1}+i \sin (\Delta t) \sigma_4^z) ]C_{34}C_{23}C_{12}	 \\
 \nonumber =& \cos(\Delta t) \mathbbm{1}^{\otimes4}  + i \sin(\Delta t)C_{12}C_{23}(\mathbbm{1}\otimes \mathbbm{1}\otimes \sigma^z_3 \otimes \sigma^z_4)C_{23} C_{12}\\
 \nonumber =& \cos(\Delta t) \mathbbm{1}^{\otimes4} + i \sin(\Delta t)C_{12}(\mathbbm{1}\otimes \sigma^z_2\otimes \sigma^z_3 \otimes \sigma^z_4)C_{12}\\
\nonumber =& \cos(\Delta t) \mathbbm{1}^{\otimes4} + i \sin(\Delta t)(\sigma^z_1\otimes \sigma^z_2\otimes \sigma^z_3 \otimes \sigma^z_4) \\
 =& e^{i\Delta \overline{Z}},
\end{align}
where $\overline{Z}=\prod_j \sigma^z_j$ is a string operator. From the above derivation, we see that the essence of the CNOT ladder is the growing of the Pauli-Z operator mediated by the nearest-neighbor CNOT gates, such that the rotation of a single qubit along z direction effectively does the exponentiation of the $Z$-string operator.   The local property in the circuit shown in Fig.~\ref{fig:circuit-transform}(a) makes it more appreciable in terms of experimental realization if only local interaction is allowed.  However, in the absence of the nearest-neighbor restriction, other  variants of this CNOT ladder exists, such as the two circuits shown in Fig.~\ref{fig:circuit-transform}(b, c).  The circuit in panel (b) contains a long-range CNOT between qubits 1 and 4, which directly adds qubit 1 onto the string started from qubit 4.  
For the circuit in panel (c), we use all long-range CNOTs between the rotated qubit 4 and other qubits, and directly mediate the Pauli-Z operator from qubit 4.  This turns out to be another extreme, which is completely non-local.  In the following, we will show a slight modification of this circuit can be transformed to our cavity-QED circuit.

\begin{figure}
  \includegraphics[width=1\columnwidth]{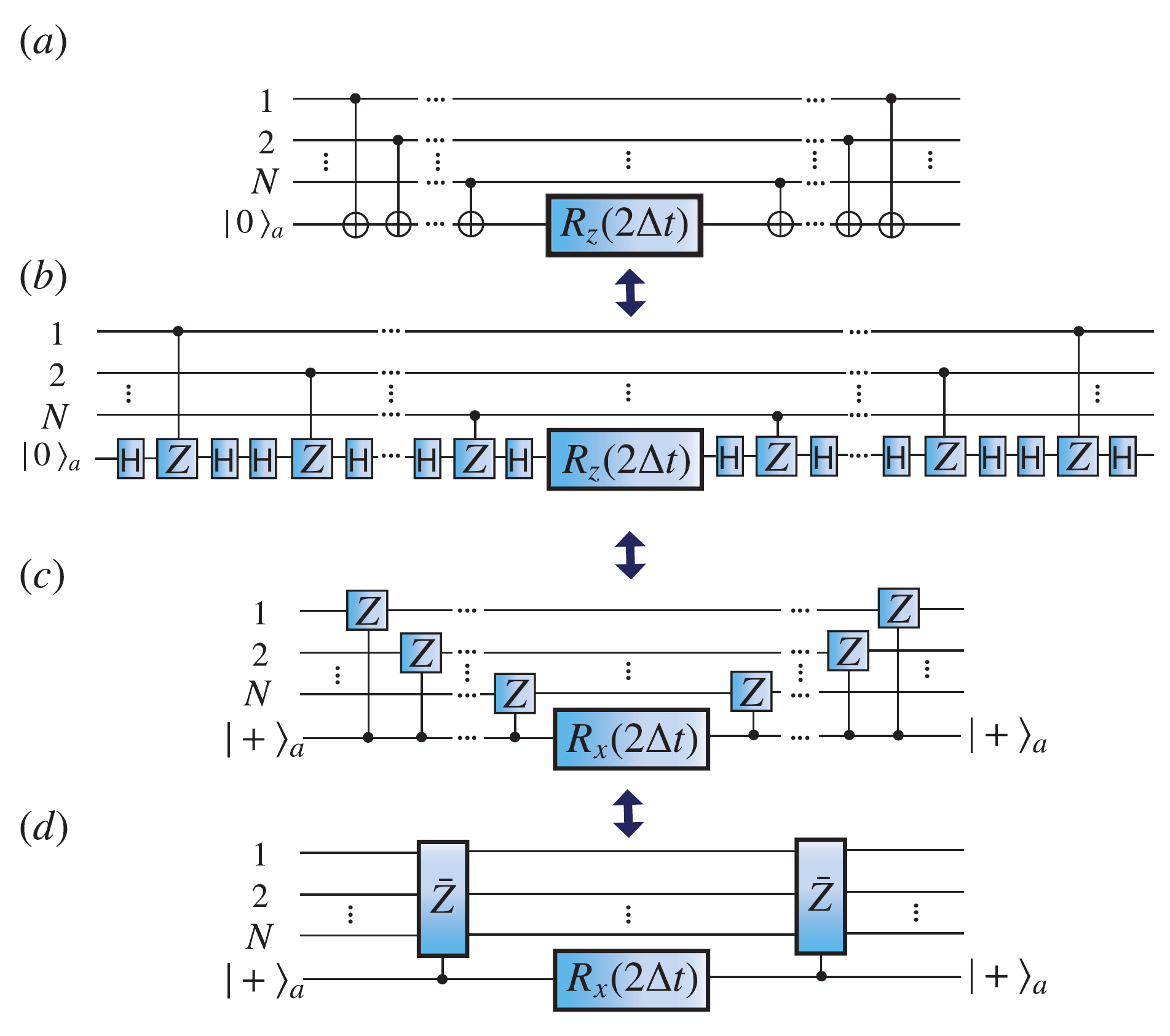}
  \caption{Circuit transformation from the CNOT ladder to the cavity-QED scheme.}
\label{fig:circuit-transform2}
\end{figure}

We consider a slight modification of the circuit in Fig.~\ref{fig:circuit-transform}(c) shown in Fig.~\ref{fig:circuit-transform2}(a) (generalized to the case of $N+1$ qubits), where now the last qubit is turned from a data qubit into ancilla and initialized into state $\ket{0}_a$ so that it does not encode fermions itself but only collects the parity information of the $N$ data qubits. By using the identity $\text{H}X\text{H}=Z$, we first transform all the CNOTs in panel (a) into CZ gates sandwiched by the Hadamard gates as shown in panel (b).  Then we use the identity $H^2=\mathbbm{1}$ to annihilate all the paired Hadamard gates except the one on the edge which transforms $\ket{0}_a$ into $\ket{+}_a$ and the two in the middle which transform the z-axis rotation $R_z(2\Delta t)$ into the x-axis rotation $R_x(2\Delta t)$.  Also we use the property that the control and target of the CZ gate is inter-changeable, i.e. CZ=ZC.  These transformations lead to the circuit in panel (c), where the controls are all moved to the ancilla.  Finally, note that, the sequential application of CZ gates can be merged into a single many-body $C_{\overline{Z}}$ gate, where $\overline{Z}=\prod_{j=1}^N \sigma^z_j$ is the string operator, as shown in panel (d), namely our cavity-QED circuit.   This is possible if the ancilla is a cavity mode which can interact non-locally with all the data qubits simultaneously and is hence hardware-efficient.  In this case, the circuit depth is reduced by a factor of $N$ due to the fact that the process of collecting the parity information into the ancilla can be done in parallel instead of the conventional approach which is in series.

\begin{figure*}
  \includegraphics[width=2\columnwidth]{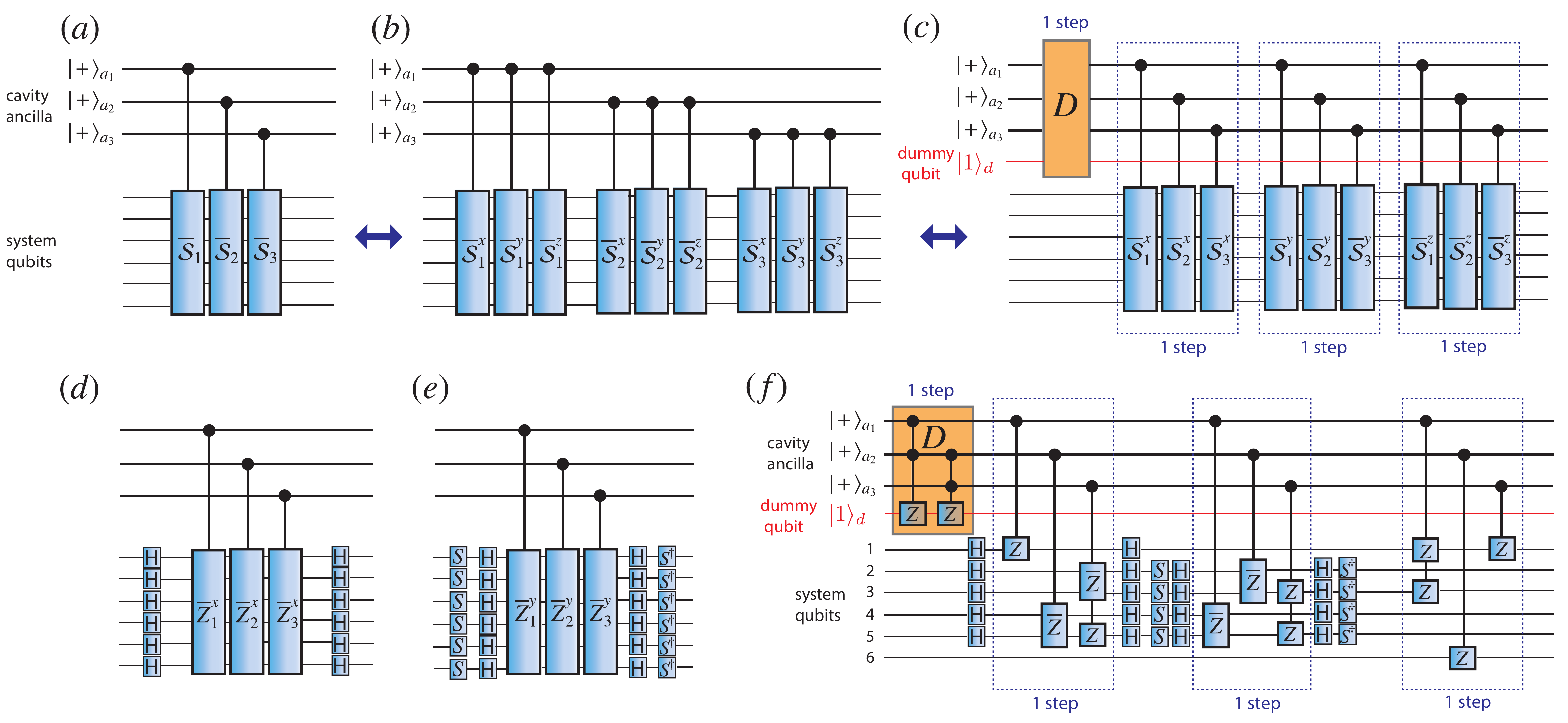}
  \caption{General strategy for parallelization multiple strings with multiple ancillae. (a) The target circuit: N arbitrary strings controlled by N ancillae.  (b) Split each string into x-, y- and z-strings.   (c) Reorder the strings such that the same type (x, y, or z) are grouped together and performed in parallel.   To fix the problem of ancillae-dependent negative sign due to anti-commutation relation between the strings, we introduce a circuit $D$ (yellow box) acting on the ancillae and a ``dummy qubit" to fix the negative sign. (d) Implementing the x-string by sandwiching the corresponding z-string with Hadmards. (e) Implementing the y-string by sandwiching the corresponding z-string with phase gates.  (f) The complete circuit for the example given by Eq.~\eqref{example}.   The circuit $D$ (yellow box) involves parallelized multi-ancillae control-Z gates acting on the dummy qubit initialized in state $\ket{1}_d$.}
\label{fig:parallelization_appendix}
\end{figure*}

\section{Multi-ancillae parallelization for a generic Hamiltonian}\label{append:general_case}

In the main text we have shown how to parallelize terms with multiple ancillae in the Fermi-Hubbard Hamiltonian, with the circuit shown in Fig.~2(f). 
Note that the ``collect" and ``erase" stages of that circuit make use of the property that the nonlocal operators of the Fermi-Hubbard model consist of long strings of $\sigma^z$-operators sandwiched between two other Pauli operators, i.e. $\sigma^x$ or $\sigma^y$. 
In this section we show how to deal with less structured Hamiltonians such as those arising from quantum chemistry in Ref.~\cite{Kandala:2017wj, Bravyi:2017wb}. 

To demonstrate the issue and our solution more clearly, we will first work out a small example explicitly and present the general treatment later. 
Consider three mutually commuting terms selected from the list of Pauli operators making up the 6-qubit Be$\text{H}_2$ Hamiltonian of Ref.~\cite{Kandala:2017wj}:
\begin{align}\label{example}
\non \overline{\mathcal{S}}_1 &= \sigma^z\sigma^x\sigma^z\sigma^y\sigma^y\mathbbm{1}, \, \overline{\mathcal{S}}_2 = \mathbbm{1}\sigma^y\sigma^y\sigma^x\sigma^x\sigma^z,\, \\
\overline{\mathcal{S}}_3 &= \sigma^z\sigma^x\sigma^x\sigma^y\sigma^x\sigma^y,
\end{align}
where we have omitted `$\otimes$' for brevity.
Here we will describe how to efficiently perform the collect/erase stage of the algorithm described in the main text, which is equivalent to the target circuit shown in Fig.~\ref{fig:parallelization_appendix}(a). 
Since we can only perform controlled $\overline{Z}$-strings using our cavity-assisted scheme [Eq.~(12) and (13)], these operators present a challenge. 
We first decompose each of the terms above into $\sigma^x$-, $\sigma^y$- and $\sigma^z$-only strings as:
\begin{align}
\nonumber
\overline{\mathcal{S}}_1^x &= \mathbbm{1}\sigma^x\mathbbm{1}\mathbbm{1}\mathbbm{1}\mathbbm{1} &	\overline{\mathcal{S}}_2^x &= \mathbbm{1}\mathbbm{1}\mathbbm{1}\sigma^x\sigma^x\mathbbm{1} &	\overline{\mathcal{S}}_3^x &= \mathbbm{1}\sigma^x\sigma^x\mathbbm{1}\sigma^x\mathbbm{1} \\
\nonumber
\overline{\mathcal{S}}_1^y &= \mathbbm{1}\mathbbm{1}\mathbbm{1}\sigma^y\sigma^y\mathbbm{1} &	\overline{\mathcal{S}}_2^y &= \mathbbm{1}\sigma^y\sigma^y\mathbbm{1}\mathbbm{1}\mathbbm{1} &	\overline{\mathcal{S}}_3^y &= \mathbbm{1}\mathbbm{1}\mathbbm{1}\sigma^y\mathbbm{1}\sigma^y \\
\overline{\mathcal{S}}_1^z &= \sigma^z\mathbbm{1}\sigma^z\mathbbm{1}\mathbbm{1}\mathbbm{1} &	\overline{\mathcal{S}}_2^z &= \mathbbm{1}\mathbbm{1}\mathbbm{1}\mathbbm{1}\mathbbm{1}\sigma^z &	\overline{\mathcal{S}}_3^z &= \sigma^z\mathbbm{1}\mathbbm{1}\mathbbm{1}\mathbbm{1}\mathbbm{1}.
\end{align}
We have shown in the main text how arbitrary controlled $\overline{\mathcal{S}}^z_\nu$-strings can be implemented using the ancilla cavity modes and arbitrary controlled $\overline{\mathcal{S}}^x_\nu$- and $\overline{\mathcal{S}}^y_\nu$-strings can be implemented by sandwiching a corresponding z-string with Hadamard gates (H) and Hadamard and phase gates ($HS$ and $HS^\dag$), respectively. 
However, as seen in Fig.~\ref{fig:parallelization_appendix}(b), the $\overline{\mathcal{S}}^x_\nu$, $\overline{\mathcal{S}}^y_\nu$ and $\overline{\mathcal{S}}^z_\nu$ strings of different terms ($\nu=1,2,3$) come in a mixed order of x-, y- and z-type, and one has to apply a pair of Hadamard/phase gates for each $\overline{\mathcal{S}}_\nu^x/\overline{\mathcal{S}}_\nu^y$, respectively. 
Moreover, all these controlled string operations cannot be parallelized.

To avoid this complication we group the controlled strings of different terms with the same type (x, y, or z) together [see Fig.\ref{fig:parallelization_appendix}(c)], by moving all the x-strings to the left and all the z-strings to the right. 
This means that given any pair of terms, the x-strings of the term to the right needs to be commuted with the y- and z-strings of the term to the left and the z-string of the term to the left needs to be commuted with the y-string of the term to the right. 
At this point we need to be careful; 
although any pair of the operators $\overline{\mathcal{S}}_1,\overline{\mathcal{S}}_2, \overline{\mathcal{S}}_3$ commute with each other, their x-, y- and z-decompositions do not have to. 
In particular, the following pairs anticommute: $\{\overline{\mathcal{S}}_1^z,\overline{\mathcal{S}}_2^y\} = \{\overline{\mathcal{S}}_1^y,\overline{\mathcal{S}}_3^x\} = \{\overline{\mathcal{S}}_1^z,\overline{\mathcal{S}}_3^x\} = \{\overline{\mathcal{S}}_2^z,\overline{\mathcal{S}}_3^y\} = 0$. 
Each of these operator swaps result in a relative minus sign which is conditioned upon the ancilla state. 
In this example [Eq.~\eqref{example}] the sign is given by
\begin{align}
	(-1)^{n_1 n_2} (-1)^{n_2 n_3} 
\end{align}
where $n_\nu$ is the photon number in ancillary cavity  mode $\nu$. 
This sign depends on the configuration of the ancilla photon numbers.
For example, in Fig.~\ref{fig:parallelization_appendix}(a, b) if the ancilla state is $\ket{101}_a$, the system is acted on by $\overline{\mathcal{S}}_1 \overline{\mathcal{S}}_3 = \overline{\mathcal{S}}_1^x \overline{\mathcal{S}}_3^x \overline{\mathcal{S}}_1^y \overline{\mathcal{S}}_3^y \overline{\mathcal{S}}_1^z \overline{\mathcal{S}}_3^z$. 
But if the ancilla state is $\ket{110}_a$, the system is acted on by $\overline{\mathcal{S}}_1 \overline{\mathcal{S}}_2 = - \overline{\mathcal{S}}_1^x \overline{\mathcal{S}}_2^x \overline{\mathcal{S}}_1^y \overline{\mathcal{S}}_2^y \overline{\mathcal{S}}_1^z \overline{\mathcal{S}}_2^z$. 
This sign can be introduced with a diagonal operator $D$ (yellow box) acting on the ancillary cavity modes and an additional dummy qubit initialized in state $\ket{1}_d$.  The detailed implementation will be discussed in the end of this section.

To be concrete,  we describe how to implement the groups of x- and y-strings. 
Given $\overline{\mathcal{S}}_i^x$ and $\overline{\mathcal{S}}_i^y$ we define corresponding z-strings by replacing all the non-identity Pauli operators with $\sigma^z$'s and call these $\overline{Z}_i^x$ and $\overline{Z}_i^y$ respectively. 
In the example considered here these are
\begin{align}
	\nonumber
	\overline{Z}_1^x &= \mathbbm{1}\sigma^z\mathbbm{1}\mathbbm{1}\mathbbm{1}\mathbbm{1} &	\overline{Z}_2^x &= \mathbbm{1}\mathbbm{1}\mathbbm{1}\sigma^z\sigma^z\mathbbm{1} &	\overline{Z}_3^x &= \mathbbm{1}\sigma^z\sigma^z\mathbbm{1}\sigma^z\mathbbm{1} \\
	\overline{Z}_1^y &= \mathbbm{1}\mathbbm{1}\mathbbm{1}\sigma^z\sigma^z\mathbbm{1} &	\overline{Z}_2^y &= \mathbbm{1}\sigma^z\sigma^z\mathbbm{1}\mathbbm{1}\mathbbm{1} &	\overline{Z}_3^y &= \mathbbm{1}\mathbbm{1}\mathbbm{1}\sigma^z\mathbbm{1}\sigma^z
\end{align} 
We then sandwich the $\overline{Z}^x_\nu$($\overline{Z}^y_\nu$)-strings with Hadamard (combined Hadamard and phase gates) to obtain the $\overline{\mathcal{S}}^x$ and $\overline{\mathcal{S}}^y$ strings, see Fig.~\ref{fig:parallelization_appendix}(d,e).

In the following, we describe the general strategy for parallelization of arbitrary terms. 
Consider a group of $N$ mutually commuting terms $\overline{\mathcal{S}}_\nu$  such that $[\overline{\mathcal{S}}_\nu, \overline{\mathcal{S}}_\mu]=0$ for $\nu, \mu=1,\dots,N$. 
In reorganizing the order of conditional strings in Fig.~\ref{fig:parallelization_appendix}(b) all the x-strings move past the y- and z-strings to its left. 
Similarly all the z-strings move past the y-strings to its right. 
A minus sign is picked up every time when the pair is anti-commuting. 
We define
\begin{align}
\tau^{\alpha \beta}_{\mu \nu} = 
	\begin{cases}
	1\, ,& \text{if } [\overline{\mathcal{S}}^\alpha_\mu, \overline{\mathcal{S}}^\beta_\nu]=0  \\
	-1\, , & \text{if } \{\overline{\mathcal{S}}^\alpha_\mu, \overline{\mathcal{S}}^\beta_\nu\}=0
	\end{cases}
	\quad	\alpha, \beta = \{x,y,z\}\, .
\end{align}
It is straightforward to show that the overall sign picked up by the rearrangement of the operators described in Fig.~\ref{fig:parallelization_appendix}(b) can be compensated by the following diagonal operator
\begin{align}
	D &= \prod_{\mu<\nu} \left(\tau_{\mu \nu}^{xy} \tau_{\mu \nu}^{xz\phantom{y}} \tau_{\mu \nu}^{yz}\right)^{n_\mu n_\nu}\, .
\end{align}
The quantity inside the parenthesis is calculated as part of the classical preprocessing. 
Whenever it is $1$, the corresponding operator is simply identity and can be discarded. 
If, on the other hand, the quantity inside the parenthesis is $-1$, we have a nontrivial contribution and the product needs to run over those pairs only:
\begin{align}
	D &= \prod_{\tau_{\mu \nu}^{xy} \tau_{\mu \nu}^{xz\phantom{y}} \tau_{\mu \nu}^{yz}=-1} (-1)^{n_\mu n_\nu}
\end{align}
Since there are $N(N-1)/2$ pairs of operators, the product above can consist of at most that many terms. 
Moreover, since a global sign is of no significance, in cases when there are more than $N(N-1)/4$ terms we can choose to implement $-D$ instead. 
With this, the worst case scenario is that the diagonal operator will consist of less than or equal to $N(N-1)/4$ terms of the form $(-1)^{n_\mu n_\nu}$. 

Next we show explicitly how to implement the sign-fixing circuit $D$ for the parallelization example with terms given by Eq.~\eqref{example} and is illustrated in Fig.~\ref{fig:parallelization_appendix}(f).  We introduce an additional ``dummy" qubit fixed in $\ket{1}_d$, and apply a control-Z gate on it when a minus sign is needed for certain ancilla configuration, due to the fact that $Z\ket{1}_d=-\ket{1}_d$.   For example, when applying a double-ancillae  control-Z gate on the  ``dummy" qubit in the case of two ancillae in total, we get
\begin{align}
\non & CCZ \bigg[(\ket{00}_a+\ket{01}_a+\ket{10}_a+\ket{11}_a) \otimes \ket{1}_d \otimes \ket{\psi}_s\bigg] \\
\non =& \bigg[(\ket{00}_a+\ket{01}_a+\ket{10}_a) \otimes \ket{1}_d \otimes \ket{\psi}_s\bigg]  \\
&- \bigg[\ket{11}_a) \otimes \ket{1}_d \otimes \ket{\psi}_s\bigg].
\end{align}
For the case in Eq.~\eqref{example} and Fig.~\ref{fig:parallelization_appendix}(f), one needs to implement an ancillae-conditioned minus sign $(-1)^{n_1 n_2} (-1)^{n_2 n_3}$, which can be realized by two double-ancilla control-Z gates (conditioned by $n_1 n_2$ and $n_2 n_3$ respectively) as shown in the yellow box in panel (f).  Note that in general all the multi-ancillae control-Z gate (more generally controlled-rotation) can be implemented in parallel utilizing the multi-mode QND interaction $H_\text{QND}'=\sum_\nu \sum_j \tilde{\chi}_{\nu,j} a^\dag_{\nu} a_{\nu} \sigma^z_j$ and the induced ``number splitting" \cite{schuster_resolving_2007}, i.e., through the different dispersive shifts of the qubit frequency depending on the number configuration in multiple cavities, i.e.  $\Delta \epsilon^j =\sum_{\nu} 2 n_{\nu} \chi_{\nu, j} $.   In the simplest two-ancillae example, we pick different dispersive shifts $\tilde{\chi}_{1,j}$ and $\tilde{\chi}_{2,j}$ due to cavity mode-1 and -2. Therefore, the total frequency shift of qubit $j$ is $\Delta \epsilon^j_{n_1, n_2}=\tilde{\chi}_{1,j}[1-(-1)^{n_1}]+\tilde{\chi}_{2,j}[1-(-1)^{n_2}]$ (with four different outcomes), where $n_1, n_2=0,1$ are the photon numbers in the two cavities. The single-qubit rotations are generated by microwave pulse with the driving frequency $\epsilon+ \Delta \epsilon^j_{n_1, n_2}$, where $\epsilon$ is the bare qubit frequency. Therefore, the driving is only resonant when the cavity-ancillae are in state $\ket{n_1 n_2}_a$. Therefore, conditional-rotations in panel (f) can be achieved. We emphasize that, in general, at most $N(N-1)/4$ conditional-rotations in the yellow box will be performed in parallel using microwave drive with multiple tones, which can be achieved with standard microwave engineering technology, e.g., for $N\sim O(50)$.

\begin{figure}
\centering
\includegraphics[width=1\columnwidth]{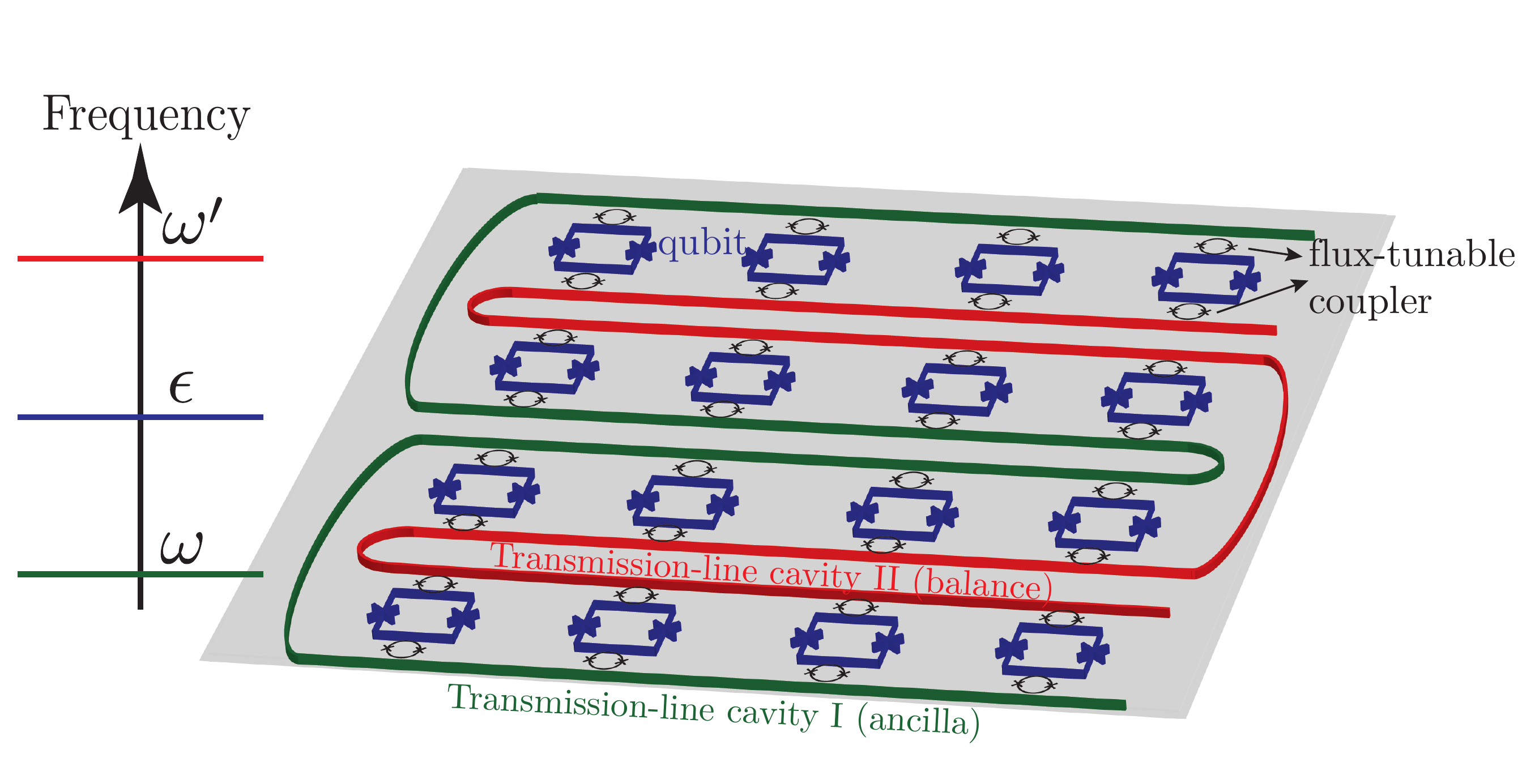}
\caption{ An on-chip circuit-QED realization with ordinary qubits and a pair of transmission-line cavities. The frequency of the qubits reside in between the frequency of the fundamental modes of the transmission-line cavities.  Superconducting qubits are coupled to two transmission-line cavities with flux-tunable inductive couplers. }
\label{fig:circuit-QED}
\end{figure}

\section{Implementation of QND interaction with the usual Jaynes-Cummings interactions}\label{append:JC}

In this section, we consider the alternative experimental realization for ordinary qubits (including transmon qubits) without the rich selection-rule structure of fluxonium.  In this case, a simple Jaynes-Cummings interaction can essentially capture the qubit-cavity coupling.

We consider the situation that all qubits are coupled to two transmission-line cavities. The qubit frequency ($\epsilon$) is placed between two different dominant cavity frequencies $\omega$ and $\omega'$ (assuming other cavity modes are far detuned away from the qubit frequency) as illustrated by Fig.~\ref{fig:circuit-QED}.  The system can be described by a two-mode Tavis-Cummings model
\begin{align}
\nonumber H=&H_0 +V  \\
\nonumber H_0=&\omega a^\dag a + \omega' b^\dag b + \frac{1}{2} \epsilon \sum_j \sigma^z_j   \\
V=&g \sum_j (\sigma^+_j a + \sigma^-_j a^\dag) +  g \sum_j (\sigma^+_j b + \sigma^-_j b^\dag).
\end{align}
Here, $a$ and $b$ represent the two photonic modes, $\sigma_j$'s represents qubits operators, and $g$ represents the strength of the Jaynes-Cummings (JC) interaction.

In the dispersive regime, namely 
\be\label{dispersive_condition2}
\sqrt{N} g \ll \abs{\Delta_a}=\abs{\epsilon-\omega} \ \text{and} \ \sqrt{N} g \ll \abs{\Delta_b}=\abs{\epsilon-\omega'}
\ee
 ($N$ represents the total number of qubits), one can adiabatically eliminate the direct JC interaction between qubits and the cavities, and the effective Hamiltonian in second-order perturbation theory is given by
\begin{align}
\nonumber H_\text{eff}=& H_0 + \frac{g^2}{\Delta_a}a^\dag a \sum_j \sigma^z_j+ \frac{g^2}{\Delta_b}b^\dag b \sum_j \sigma^z_j  \\
&+ \bigg(\frac{g^2}{\Delta_a}+ \frac{g^2}{\Delta_b} \bigg) \sum_{j<j'}(\sigma^+_j \sigma^-_{j'} +\text{H.c.})+\frac{g^2}{2\Delta}\sum_j \sigma^z_j+O(g^4).
\end{align}
Apart from $H_0$, the terms appearing in second-order perturbation have two types: (1) The QND interaction between the cavity photons and the qubits [Eq.~(3)];  (2) The non-local flip-flop interactions between qubits mediated by virtual photons.  For the purpose of our protocol, we want to get rid of the later type.  This can be simply achieved by setting $\Delta_a=-\Delta_b$, i.e.~placing the qubit frequency right in the middle of two cavity frequencies [$\epsilon=\frac{1}{2} (\omega+\omega')$],  as illustrated by Fig.~\ref{fig:circuit-QED}. We call the first transmission line ``ancilla cavity", and we occupy this cavity (mode $a$) with photons. We call the second transmission line  ``balance cavity".  We will not occupy this cavity (mode $b$) with photons, and one can effectively set $b^\dag b=0$. 

For the above discussion,  the cancellation of non-local flip-flop terms relies on uniform qubit-cavity coupling $g$.  However, usually $g$ can have spatial dependence due to non-uniform shape of the mode function.  Therefore, one should also be able to vary the qubit-resonator coupling strength on different sites to compensate such inhomogeneity.    This can be achieved with the flux-tunable inductive couplers between qubits and transmission-line cavities as illustrated in Fig.~\ref{fig:circuit-QED}.

Now, the only remaining term apart from $H_0$ is the QND interaction between the ancilla photon and the qubits namely $H_\text{QND} = \chi a^\dag a \sum_j \sigma^z_j$,
where the interaction strength is $\chi= g^2 / \Delta_a$.   

Note that an alternative way of suppressing the non-local filp-flop interactions without using a balance cavity is by detuning the frequencies of the qubits which are coupled to the same cavity mode. As long as the frequency difference $\Delta \epsilon$ is much larger than the QND interaction strength $\chi$, namely $\Delta \epsilon /\chi \ll 1$, the flip-flop interaction is effectively suppressed due to rotating-wave approximation.    This alternative scheme works well for $N\sim O(10)$ since the QND interaction is still sizeable.

\section{Fluxonium circuit}\label{append:fluxonium}

\begin{figure}[hbt]
  \includegraphics[width=1\columnwidth]{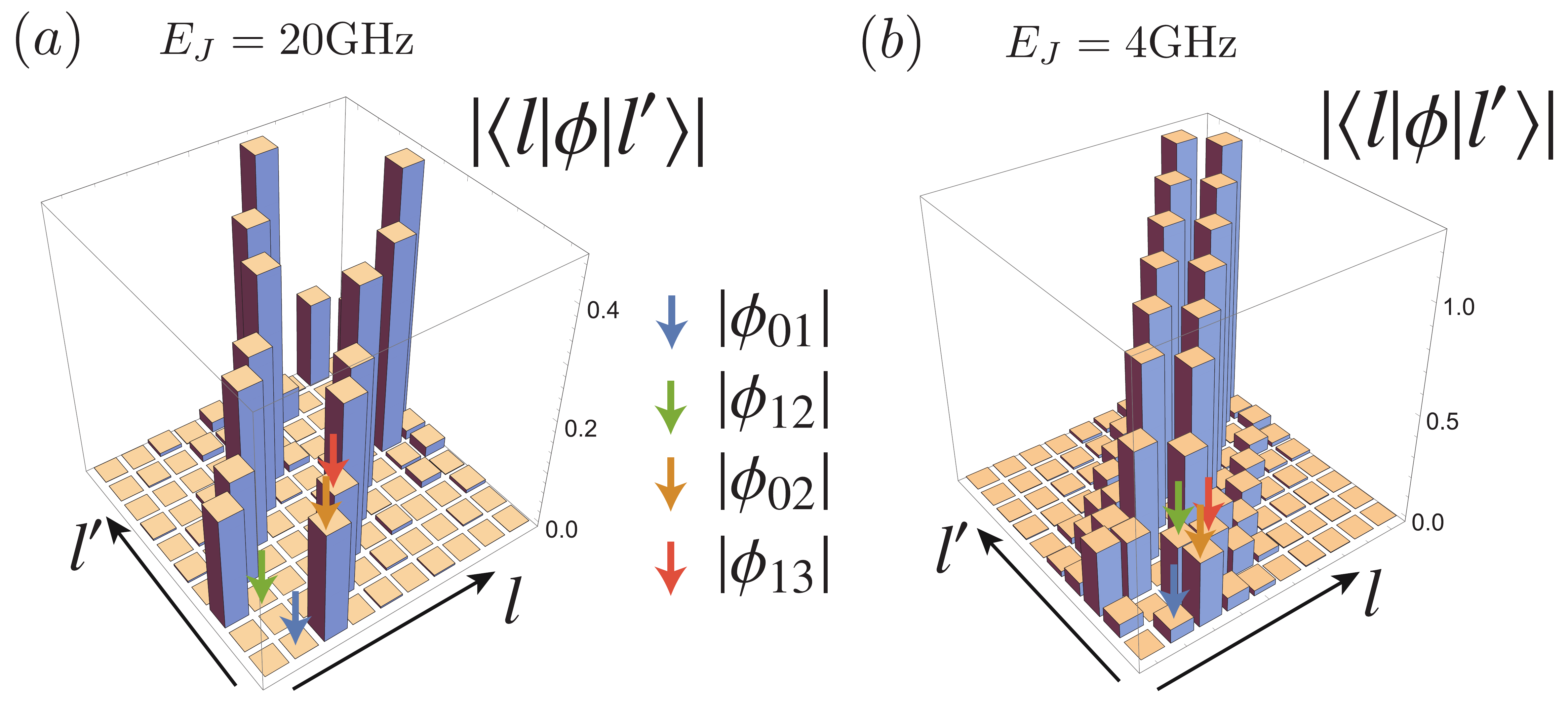}
  \caption{Phase matrix elements of the fluxonium in the large (20 GHz) and small (4 GHz) $E_J$ regime, with the other parameters fixed: $E_C= 0.5 \text{GHz}$ and $E_L= 0.75 \text{GHz}$.   }
  \label{fig:matrix_elements}
\end{figure}

The fluxonium circuit can be described the following Hamiltonian \cite{manucharyan_fluxonium:_2009, Zhu:2013kf, Zhu:2013fa}: 
\be
H_f = 4 E_C N^2 - E_J \cos \phi + \frac{1}{2} E_L (\phi + 2\pi \Phi_\text{ext}/\Phi_0)^2,
\ee
where $\phi$ describes the phase difference across the small junction, the conjugate operator $N = -i d/d\phi$ represents charge imbalance across the junction, in units of Cooper pair charge (2e),  and $\Phi_\text{ext}$ represents the flux threading the main loop of the circuit.  The relevant energy scales are charging energy $E_C$ and Josephson energy $E_J$ of the small junction, and the effective inductive energy $E_L$  of the ``superinductor" formed by a Josephson junction array.  One can think of the above Hamiltonian as describing a fictitious particle (with coordinate $\phi$) residing in a potential $V(\phi)=- E_J \cos \phi + \frac{1}{2} E_L (\phi + 2\pi \Phi_\text{ext}/\Phi_0)^2$, which is composed by a periodic cosine potential and an parabolic envelope (position tunable with the external flux $\Phi_\text{ext}$). The charging term is equivalent to the kinetic energy of the particle with momentum being $N = -i d/d\phi$.  One can define an orthogonal Wannier basis $\ket{m}_s$ according the first two terms,  i.e.~particle in a cosine periodical potential.  Here, $m$ labels the wells in the cosine potential and also represents the winding number of the persistent current corresponding to the particular Wannier state, while $s$ is the band index.   In this basis,  one can write the effective Hamiltonian for each band $s$ (neglect inter-band coupling) as follows:
\begin{align}
\non H_s \approx & \frac{(2\pi)^2}{2} E_L (m + \Phi_\text{ext}/\Phi_0)^2   \\
&+ \frac{1}{2} \sum_{m=-\infty}^{\infty} \epsilon_{s,1}[{_s}\ket{m}\bra{m+1}_s + \text{H.c.}] + \epsilon_{s,0},
\end{align}
where $ \epsilon_{s,n}$ is defined by the Fourier expansion of the band structure $\epsilon_s(p) = \sum_n \epsilon_{s,n}\cos(2\pi n p)$ ($p$ is the quasi-momentum).    We see that once the leading-order inter-well tunneling amplitude $\epsilon_{s,1}$ is suppressed (in the large $E_J/E_C$ regime),  the eigenstate of $H_s$ are approximately Wannier states $\ket{m}_s$ in different wells ($m$) and bands ($s$).  Therefore, $m$ effectively becomes a good quantum number, and hence one obtains a selection rule ${_s}\boket{m}{\phi}{m'}_{s'}\propto \delta_{mm'}$ \cite{Zhu:2013fa}, i.e.~inter-well transitions are forbidden due to the exponential suppression of wavefunction overlap.  The phase matrix elements $\boket{l}{\phi}{l'}$ are illustrated for large and small $E_J$ in Fig.~\ref{fig:matrix_elements}.  At large $E_J$ [panel (a)] one can see clear selection rule, i.e. inter-well transitions are forbidden ($\phi_{01} =\phi_{12}=0$), while at small $E_J$ [panel (b)], no clear selection rule exists and the matrix elements have a richer structure.

An alternative scheme of cancelling the flip-flop interaction between fluxonium qubits without using selection rules in the specific regime is to spatially vary the parameters (such as $\Phi_\text{ext}$ or $E_J$) such that the 0-1 transition energy is detuned between different qubits.    Due to the rich structure of the matrix elements, the dispersive shifts $\chi_l = \sum_{l'} \chi_{ll'} $ and QND interaction strength $\chi=\frac{1}{2}(\chi_1 -\chi_0)$ receive contributions from multiple levels $l'$ and hence usually does not decreases when the 0-1 detuning $\Delta_{01}$ increases as in the case of usual Jaynes-Cummings model (for two-level systems).    Ultimately frequency crowding limits the number of qubits allowed in this scheme.   

\section{Circuit design for Fermi-Hubbard model}\label{append:Hubbard}

\begin{figure}
  \includegraphics[width=1\columnwidth]{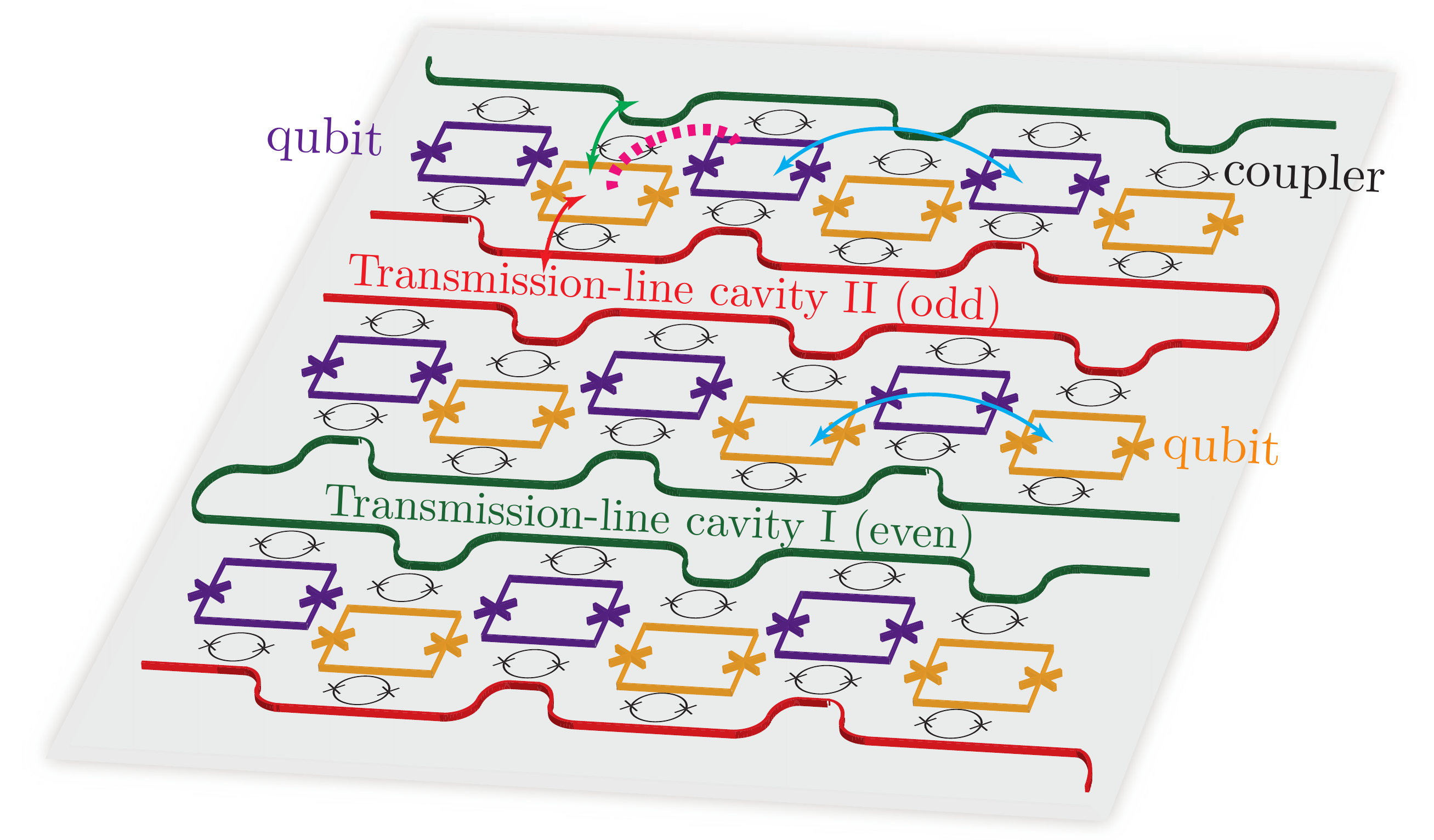}
  \caption{Experimental implementation with multiple transmission-line cavities that can parallelize the exponentiation of sub-terms in each pair of neighboring rows.  The flux couplers can be turned on and off such that they can be used to couple qubits to exponentiate two-body terms or be used to couple qubit and cavity to exponentiate terms with strings.}
\label{fig:Hubbard_circuit}
\end{figure}

In this section we discuss the on-chip circuit-QED design for the implementation of the Fermi-Hubbard model.  For the purpose of parallelization,  we couple each neighboring pair of rows with a cavity through a tunable flux couplers.
Note that each qubit is coupled to two cavities (red and green as shown in Fig.~\ref{fig:Hubbard_circuit}).  In the following, we discuss the three types of terms mentioned in the main text  with more details.

(1) \textit{On-site Hubbard interaction}:  
This interaction translates to ZZ interaction and phase shift in the qubit representation, i.e.~
\[
U \sum_j  n_{j,\uparrow} n_{j, \downarrow} \longrightarrow  U \sum_j (2\sigma^z_{j, \uparrow}-1)(2\sigma^z_{j, \downarrow}-1).
\]
While the phase shift can be easily implemented by shifting the qubit frequency, the $ZZ$ interaction can be induced perturbatively (second-order) by the flux-tunable inductive couplers between the qubits of two spin species  (purple and yellow) [illustrated by the red dashed line in Fig.~\ref{fig:Hubbard_circuit}].

(2) \textit{Horizontal hopping}: as illustrated in Fig.~4 in the main text, horizontal hopping (for both spin species) does not contain a JW string, and hence can be directly implemented by the inductive coupler between neighboring qubits [illustrated by the blue double arrows in Fig.~\ref{fig:Hubbard_circuit}].   

(3) \textit{Vertical hopping (even and odd)}:  The vertical hopping between neighboring rows typically contains a JW string and can be exponentiated with the circuit in Fig.~2(b) in the main text. The hopping terms are divided into two generally non-commuting groups for both spin species, with the lower row having even (odd) row index, with the corresponding JW string on the left (right) side due to the ``snake"-shape of the JW string in our convention, as illustrated in Fig.~4 in the main text.
To exponentiate these two groups of terms in turn, we couple the qubits on the corresponding two neighboring rows through inductive couplers with a green (red) transmission-line cavity for the even (odd) vertical hoppings. We also exponentiate the two spin species in turn.  The inductive couplers in this design can be used to couple two qubits, or a qubit with the resonator.   

During the protocol of exponentiating the above types of terms respectively, one can turn the couplers on and off and detune the qubits properly, such that different processes do not interfere with each other.

\section{Generalization to Bravyi-Kitaev transformation and Fenwick tree encoding}\label{append:bravyi}
In addition to the Jordan-Wigner encoding introduced earlier, there are a number of other fermionic encoding schemes. 
The fermionic antisymmetric space is spanned by basis states $\ket{K}=\prod_{i=1}^M (c_i^{k_i})^\dag\ket{\Omega}$, where $\ket{\Omega}$ is the vacuum.
The action on state $K$ of the lowering operators is given by $c_j\ket{K}=(-1)^{\Gamma_{jK}}\ket{K'}\delta_{k_j, 1}$ with $K'$ such that $k_i'=k_i$ for $i\neq j$ and $k'_j=0$.  Thus, changing the occupancy requires knowledge of the prefix sum $\Gamma_{jK}=\sum_{i=1}^{j-1} k_i$ and the occupancy $k_j$.  Depending on the encoding scheme, either the computation of the occupancy or of the prefix sum may cause the qubit operator locality to grow with system size.  The simplest case is the JW transform where only the occupancy is stored in the qubits.

The optimal compromise between the two schemes results in Fenwick tree where a mixture of occupancy and partial sums are stored, known as the Bravyi-Kitaev (BK) transform.   Despite additional complications in prescribing operators, the total cost (qubit string length) to encode a fermionic mode is $\log N$ for $N$ fermionic modes.  
Here, we show a concrete example of exponentiating the qubit string in BK encoding in a system with six fermionic modes.   In particular, we consider the encoding of
a hopping term between the first and last modes, i.e.~$h_{16}=\kappa c_1^\dag c_6 + \text{H.c.}=\frac{\kappa}{2} (\sigma^x_1 \sigma^x_2 \sigma^x_3 \sigma^z_6 + \sigma^y_1 \sigma^x_2 \sigma^y_3 \sigma^z_5 )$.  We present the corresponding circuit for the second sub-term in Fig.~2(d) in the main text.

\section{Measurement of the dynamical correlation function}\label{append:dynamical_correlator}

\begin{figure}
  \includegraphics[width=1\columnwidth]{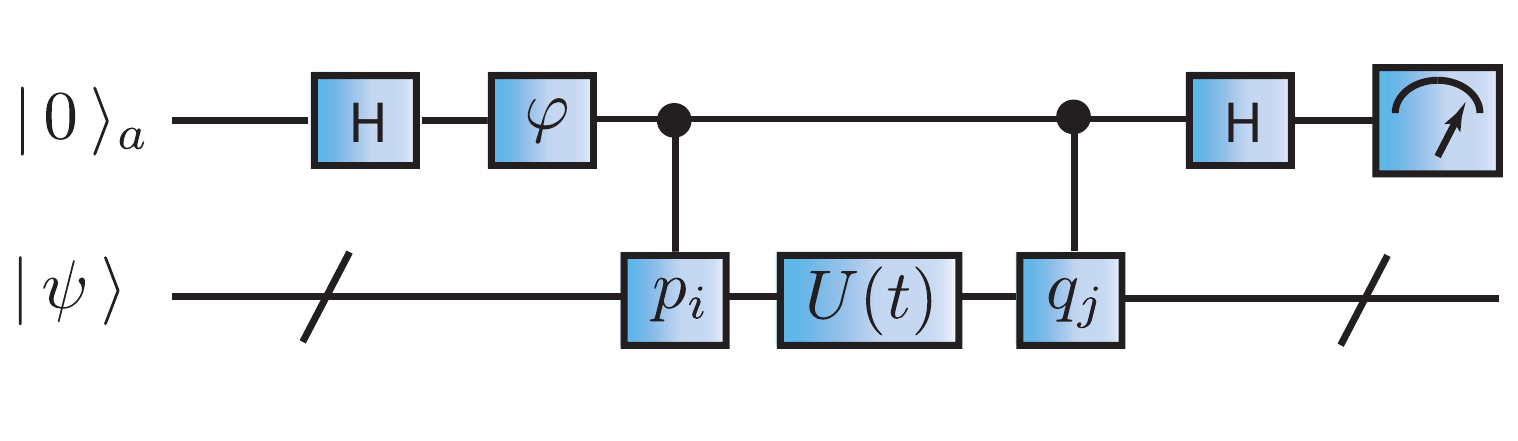}
  \caption{Ramsey interference circuit for measuring the dynamical correlator $\boket{\psi}{q_i(t) p_j(0)}{\psi} $.}
\label{fig:dynamical_correlator}
\end{figure}

A useful measure which captures the response of the system is the dynamical correlator:  
\be
G^p_{ij}(t)=\boket{\psi}{c_i(t) c^\dag_j(0)}{\psi}; \ G^h_{ij}(t)=\boket{\psi}{c^\dag_i(t) c_j(0)}{\psi},
\ee
where $G^p$ and $G^h$ represents particle and hole correlators respectively.  In order to make the measurement easier, we consider measuring correlation functions of the related Hermitian operators, namely the Majorana operators defined as:
\be
q_j = c_j+c^\dag_j, \quad p_j=i(c_j-c^\dag_j).
\ee
One can express the complex fermion correlator as linear combinations of Majorana correlators as
\begin{align}\label{Majorana_definition}
\nonumber G^p_{ij}(t)=&(\boket{\psi}{q_i(t) q_j(0)}{\psi} + i \boket{\psi}{q_i(t) p_j(0)}{\psi})/2, \\
G^h_{ij}(t)=&(\boket{\psi}{q_i(t) q_j(0)}{\psi} - i \boket{\psi}{q_i(t) p_j(0)}{\psi})/2.
\end{align}
Note that since we have fermion number conservations in the Fermi-Hubbard model, the other two types of Majorana correlators $\langle pp \rangle$ and $\langle pq \rangle$ does not appear in the above interaction, although they should appear if there are number-non-conserving terms in the Hamiltonian.  Therefore we only need to measure the two types of Majorana correlators, i.e.~$\langle qq \rangle$ and $\langle qp \rangle$.  Using $\langle qp \rangle$ as an example, we can re-express it in the Schr\"odinger picture using evolution operator $U(t)$, i.e.~
\be
\boket{\psi}{q_i(t) p_j(0)}{\psi} = \boket{\psi}{ U^\dag(t) q_i U(t) p_j}{\psi}.
\ee
One can easily see that this can be easily measured by taking an overlap between $ q_i U(t) p_j\ket{\psi}$ and $ U(t)\ket{\psi}$.  Therefore, one can again use a Ramsey interference protocol to extract the correlator from the ancilla (cavity), which is shown in Fig.~\ref{fig:dynamical_correlator}.  The only needed ingredients are unitary evolution and controlled Majorana operator.  The latter can again be realized with the combination of controlled string operation $CZ$ [Eq.~(5)], Hadamard and phase gates.  When setting the phase rotation of the ancilla $\varphi=0$ ($\varphi=\pi/2$), one get the real (imaginary) part of the correlator. 

Once the Majorana correlators are measured and one converts them into the particle and hole correlators, one can now calculate the spectral function from a Fourier transform of the dynamical correlator on the same site ($i=j$):
\begin{align}\label{spectral_function}
\non G(\omega) =& -i\left[\int_{\epsilon}^{\infty}dt e^{i(\omega+i\eta)t} G^p_{ii}(t) +  \int_{\epsilon}^{\infty}dt e^{-i(\omega-i\eta)t} G^h_{ii}(t)   \right]\\
A(\omega) =& -2\text{Im}G(\omega),
\end{align}
where $A(\omega)$ is the spectral function and encodes important information of the system (such as the spectral gap).  Note that $\epsilon$ and $\eta$ are infinitesimal real number to shift the poles and integration contour.

\begin{figure*}
\centering
\includegraphics[width=1.8\columnwidth]{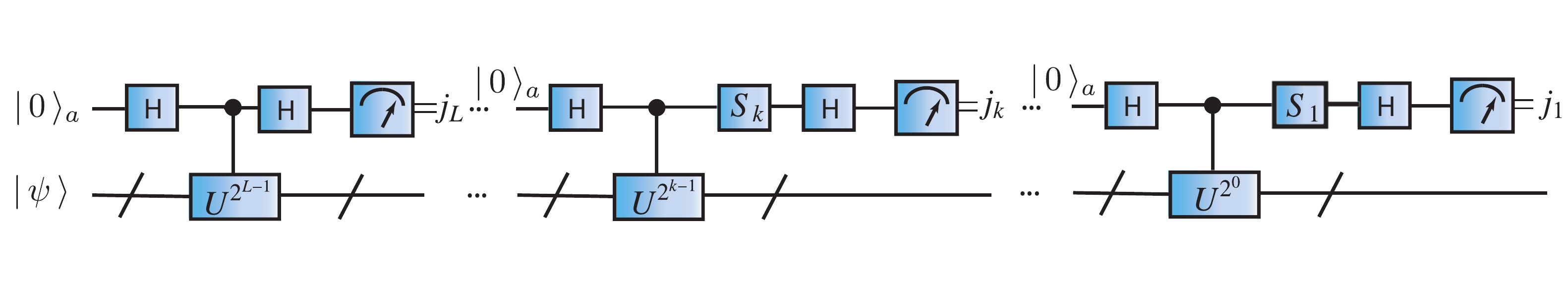}
\caption{Iterative phase estimation (IPEA) algorithms to measure the spectrum and prepare eigenstates or more specifically the ground state of the Hamiltonian.}
\label{fig:phase_estimation}
\end{figure*}

\section{Phase estimation (PEA) of the energy spectrum and state preparation}\label{append:phase_estimation}

In this section, we discuss the implementation of both the `analog' Kitaev-Ramsey phase estimation used to extract the energy spectrum, and the `digital' quantum phase estimation \cite{Kitaev:1995tq, nielsen2000quantum} which can also be used for state preparation.

For the former, one starts with the ancilla in state $\ket{0}_a = \frac{1}{\sqrt{2}} (\ket{+}_a+\ket{-}_a)$ and many-body state $\ket{\psi}_0 =\sum_k C_k \ket{E_k}$ (represented by eigenstates with energy $E_k$). Following Eq.~(7) and Fig.~2(a) in the main text, one then applies a conditional evolution 
\be\label{condtional_evolution}
CU=e^{-iHt} \ketbra{+}_a + e^{iHt} \ketbra{-}_a, 
\ee
where $t=n\Delta t$. The final state has the entangled form, i.e.~$\ket{\psi}_f$$=$$ \sum_k  \frac{C_k}{\sqrt{2}} ( e^{-iE_kt} \ket{E_k} \otimes\ket{+}_a + e^{iE_kt} \ket{E_k} \otimes \ket{-}_a)$. 
Therefore, the measurement of the ancilla in Z-basis gives rise to 
\be\label{ancilla_estimation}
\text{Re}[\langle Z_a (t)\rangle] = \sum_k \abs{C_k}^2 \cos (2E_k t),
\ee
from which the energy spectrum, $\{E_k\}$ can be inferred from the Fourier transform of the above time-domain signal into frequency-domain, i.e., $\text{Re}[\langle Z_a (\omega)\rangle]$, while the weight of the peaks are determined by the weights $C_k$ in the initial state $\ket{\psi}_0$. This is nothing but the Kitaev-Ramsey phase estimation algorithm \cite{Kitaev:1995tq, nielsen2000quantum}. The ancilla measurement $\langle Z_a (t)\rangle$ and $\text{Re}[\langle Z_a (\omega)\rangle]$ are the quantities simulated in Fig.~5 in the main text.

When using the property of Eq.~\eqref{condtional_evolution} to perform PEA, one needs to exponentiate each sub-term of the Hamiltonian with a single cavity ancilla sequentially, so the parallelization scheme in the previous section cannot be used anymore.   In the case of a 2D spinful Fermi-Hubbard model in real space ($N \times N$ lattice),  the circuit-depth (time-complexity) of a single Trotter cycle in the phase estimation becomes $O(N^2)$ if Jordan-Wigner encoding is used. In the case of the generic Hubbard model with $N$ orbitals, it remains $O(N^4)$, the same as the circuit in series as we discussed above.   In order to use the parallelization scheme,  we can use an alternative approach, namely coupling the multiple ancilla cavities ($a_j$) to a central ``clock" cavity ($a_c$), which shifts the frequencies of the other cavities conditionally through the dispersive coupling ($\chi a^\dag_j a_j a_c^\dag a_c$) such that the rotations applied on all the ancilla cavities in parallel is conditioned by the state of the clock cavity.  In this scheme, the time complexity of PEA remains the same as ordinary time evolution.  

In contrast, note that in the conventional approach, the phase estimation assumes the coupling to a central ancilla, which is very inefficient to implement with an experimental system involving only local interactions.   In that case, one has to apply an extensive amount of SWAP gates to transport the ancilla around.  Therefore, having an cavity ancilla coupled non-locally to the whole system has a significant advantage for phase estimation.

\begin{figure*}[hbt]
\centering
\includegraphics[width=1.5\columnwidth]{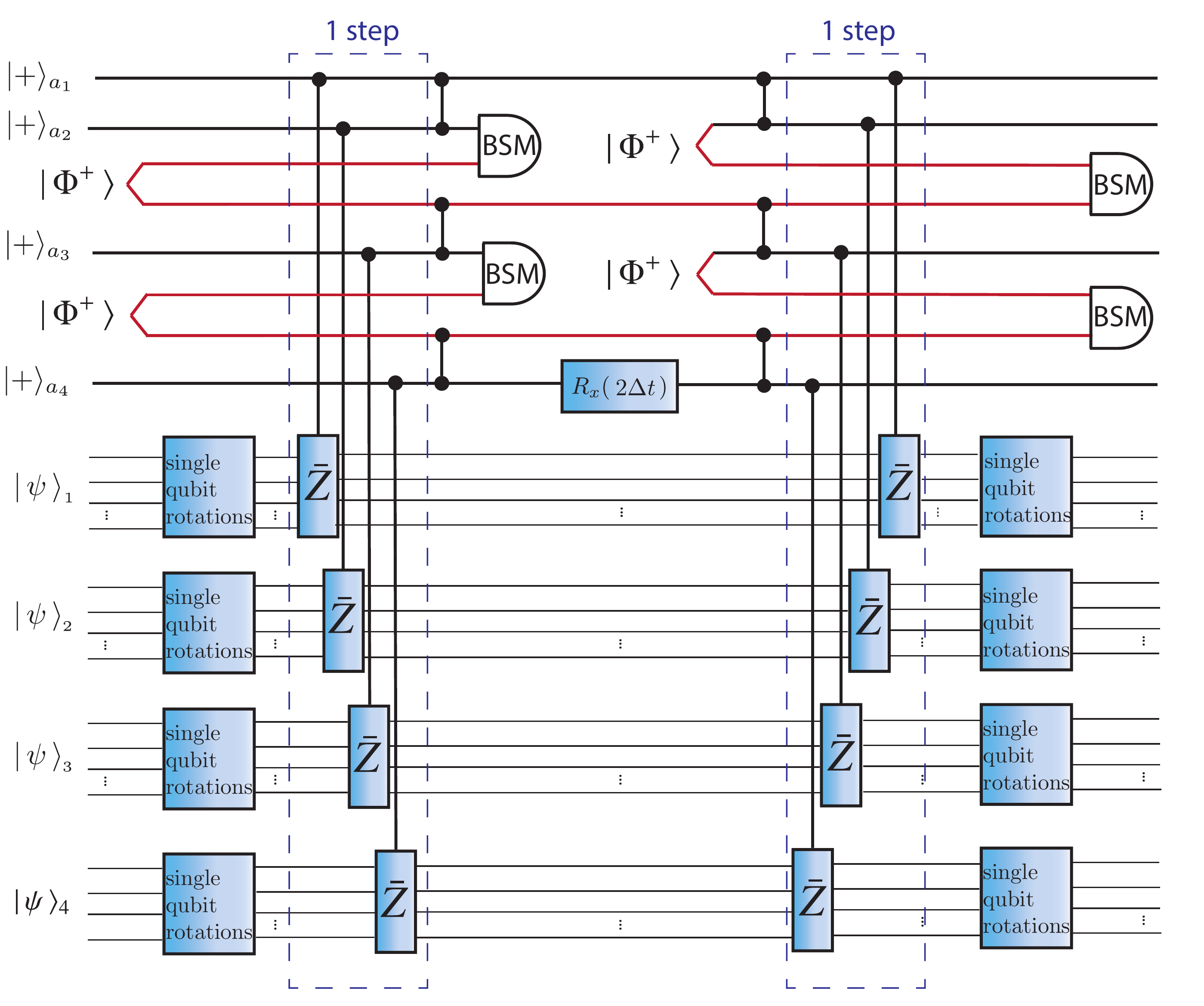}
\caption{Scalable architecture with coupled cavity modules  that stitch qubit strings in each module together. Four modules are illustrated here, and each module contains multiple qubits and one cavity mode which can be easily generalized to the multi-mode case. The red wires represent the additional ancilla qubits enabling teleportation in the spirit of Ref.~\cite{Raeisi:2012hc}.  The ancilla are initially prepared in Bell state $\ket{\Phi^+} = \frac{1}{\sqrt{2}} (\ket{00}+\ket{11})$ and are measured (and corrected) jointly with the neighboring cavity in the Bell-state basis to teleport the parity information the cavities have collected to the target cavity $a_4$. The target cavity is rotated in order to  exponentiate the qubit strings. After that,  a mirror teleportation circuit is applied to erase the parity information.  The circuit depth is of $O(1)$. }
\label{fig:scalable}
\end{figure*}

Now we consider the `digital' version of quantum phase estimation \cite{nielsen2000quantum}, the essence of which can be summarized as
\begin{align}
\nonumber \frac{1}{\sqrt{N}}\sum_t \ket{t}\otimes\ket{\psi} &\xrightarrow{CU} \frac{1}{\sqrt{N}} \sum_{k,t} C_k e^{-iE_k t} \ket{t} \otimes \ket{e_k} \\
& \xrightarrow{QFT}\sum_k C_k \ket{E_k} \otimes \ket{e_k}.
\end{align}
Here, the first register is called the ``time register", which is initialized in superposition of different ``time", i.e. $\frac{1}{\sqrt{N}}\sum_{t=0}^{2^N-1}\ket{t}$, where $N$ is the number of qubits in the time register. The second register is called the ``state register" containing the many-body wavefunction $\ket{\psi}$.  Now one applies a controlled unitary $CU$ for time $t$ determined by the time register state $\ket{t}$.  The relative phase factor $e^{-iE_k t}$ is hence imprinted into the entangled state of $\ket{t}$ and the energy eigenstate $\ket{e_k}$.   Here, $C_k = \bket{e_k}{\psi}$ is the amplitude of the many-body wavefunction being in eigenstate $\ket{e_k}$.  One can further perform a quantum Fourier transform (QFT) such that the time register now stores the estimated energy $E_k$ of the corresponding many-body energy eigenstate $\ket{e_k}$.  By doing projective measurement on the $N$ qubits in the time register, one can get the estimated energy $E_k$ and meanwhile project the many-body state into the energy eigenstate $\ket{e_k}$ with probability $\abs{C_k}^2$, which can also be used as a state preparation protocol.    

 In the context of our cavity-QED scheme, there is only one ancilla qubit in the time register, which is the cavity photon state.  Therefore,  we use the iterative phase estimation algorithm (IPEA) \cite{Parker:2000dj} to repetitively use the single ancilla qubit to improve the precision of energy measurement and state preparation, the quantum circuit of which is shown in Fig.~\ref{fig:phase_estimation}.   In the PEA language, one can represent the phase factor generated by the unitary evolution $U(t_0)=\exp(-iHt_0)$ as $\exp(-i E t_0)=\exp(-2\pi \phi)$ where $\phi \in [0,1)$ is the phase to be estimated.  The phase $\phi$ can be represented with a fractional binary expansion
 \be
 \phi=0.j_1 j_2 ... j_L =\frac{j_1}{2^1} +\frac{j_2}{2^2} + \cdots + \frac{j_L}{2^L}.
 \ee
 As the first step in IPEA, a conditional $U(2^{L-1}t_0)$ is performed (abbreviated as $U^{2^{L-1}}$), resulting in the relative phase factor $\exp[2\pi i ( j_1 ... j_{L-1}. j_L)]=\exp(-2\pi i j_L/2)$. Therefore, the least-important final digit $j_L=0$ or $1$ can be measured through the readout of the cavity ancilla.  The measurement projects the many-body wavefunction in the state register and this output wavefunction in the first step automatically serves as the input of the next iterative step.   Now to obtain more significant digits $j_k$, we need to perform a conditional $U(2^{k-1} t_0)$ (abbreviated as $U^{2^{k-1}}$), as shown in Fig.~\ref{fig:phase_estimation}. In this case, we get the phase factor $\exp[-2\pi i ( j_1\cdots j_{k-1}.j_{k}\cdots j_L)]$. We see that the more significant digits on the left decimal point does not contribute to the phase factor in this case.  However, we need to get rid of the less significant digits on the right of $j_k$ to make the measurement precise by the additional phase gate $S_k$ applied on the ancilla before the measurement.  The form of $S_k$ is given by 
\be
S_k= \begin{pmatrix}
 1 & 0  \\
 0 & \exp\left[ 2\pi i \sum_{l=2}^{L-k+1} \frac{j_{k+l-1}}{2^l} \right], 
\end{pmatrix}
\ee    
and is based on the previous measurement results of the less significant digits $j_{k+1}, j_{k+2} \cdots j_L$. With this phase gate, one ends up with the phase factor $\exp[-2\pi i  0.j_{k}] = \exp[-2\pi i j_{k}/2]$.  The measurement of the ancilla hence gives the digit $j_k$.  By iterative performing such measurement procedure, one gets all the digits and hence the estimated phase $\phi=0.j_1 j_2 ... j_L$.  Meanwhile, the many-body state in the state register is projected to the eigenstate corresponding to the eigenenergy $2\pi \phi$.  Therefore, IPEA can be used to prepare the eigenstate or more specifically the ground state of the system with order of time $O(1/\epsilon)$, where $\epsilon$ is the precision of the eigenenergy.  For the ground state, the preparation time is of the order $O(1/\Delta_M)$, where $\Delta_M$ is the many-body gap.  This is in contrast to an alternative way of state preparation, i.e.~adiabatic state preparation, which requires time of the order $O(1/\Delta_M^2)$.

	



\section{Scalable modular architecture and stitching of the strings}\label{append:modular}
Due to the constraint of the qubit number in one cavity limited by experimental feasibility and also the asymptotic $1/\sqrt{N}$ (or $1/\sqrt{\log(N)}$ for BK encoding) reduction  of the allowed perturbative interaction strength, it is important to consider scalable architecture that couples multiple modules together, where each module consists of a cavity with multiple modes and a number of qubits.  We show such a scalable scheme in Fig.~\ref{fig:scalable}, which manages to stitch the qubit strings in multiple modules together. For simplicity we only show a single cavity mode in each module, while it can be extended to the situations with multiple modes in each module.   Our modular scheme uses a teleportation scheme which can be considered as a variant of the teleportation circuit in Ref.~\cite{Raeisi:2012hc}. 

In our scheme, the cavity modes collects the parity information of the qubit strings within each module and is teleported to a single target cavity with the assistance of additional ancilla qubit pairs (red lines) which are prepared in Bell state $\ket{\Phi^+} = \frac{1}{\sqrt{2}} (\ket{00}+\ket{11})$.  The teleportation is completed by the Bell-state measurement (BSM) projection and conditional correction again to the target state $\ket{\Phi^+}$.  The target cavity is rotated by $R_x (2\Delta t)$ in order to exponentiate the string operator.   After that, a mirror teleportation circuit is implemented to erase the parity information.   Note that, since all the Bell-state preparation and measurement, and the conditional string operations in each module are performed in parallel, the circuit depth of the modular architecture to exponentiate a single string is $O(1)$.  Therefore modular approach does not increase the time complexity and can remedy the slowing down of the cavity-QED scheme due to the asymptotic $1/\sqrt{N}$ reduction in interaction strength.

\section{Discussion and comparison with other relevant schemes}\label{append:discussion}
In this section, we compare our cavity-QED scheme to the schemes in Ref.~\cite{Raeisi:2012hc, Hastings:2014uj}.    

As pointed out by Ref.~\cite{Hastings:2014uj}, one can 
also reduce the Trotter step circuit depth by a factor of $O(N)$ asymptotically if using a non-local ancilla coupling to all the qubits to enable certain cancellation.
Note that this non-local ancilla can be naturally implemented by the cavity we propose in this paper.   Also, if using our cavity-assisted string operation, one can get a further constant reduction comparing to the scheme in Ref.~\cite{Hastings:2014uj} since the leftover CNOT ladder can be merged into a single $C\overline{Z}$ gate.  On top of this cancellation, Ref.~\cite{Hastings:2014uj} points out that one can parallelize certain ``nesting" terms (with multiple strings nested in a parent string) and get another $O(1)$ reduction, while our parallelization scheme using multiple ancillae is more general and does not rely on such nesting.

In Ref.~\cite{Raeisi:2012hc}, a teleportation circuit was proposed to replace the CNOT ladder, which uses Bell-state measurement between every two neighboring qubits, hence one can also get an $O(N)$ reduction in circuit depth.   Note the measurement fidelity is typically lower than the gate fidelity in experimental platforms such as superconducting qubits. Moreover, our cavity scheme has the advantage of parallelization with multiple ancillae.  However, it is worth mentioning that the teleportation approach can be incorporated into our modular architecture connecting the strings in multiple cavities as discussed in Appendix \ref{append:modular}.

\end{appendix}


\end{document}